\documentclass[a4paper,fleqn,usenatbib,useAMS]{mnras}

\usepackage{graphicx}	
\usepackage{amsmath}	
\usepackage{amssymb}	
\usepackage{multicol}        
\usepackage{bm}		
\usepackage{pdflscape}	
\usepackage{comment}

\newcommand{\kms}{{\hbox{km s$^{-1}$}}} 

\newcommand{\etal}{et~al.}
\newcommand{\ebv}{\hbox{$E(B-V)$}}
\newcommand{\Msol}{M$_\odot$}

\usepackage[T1]{fontenc}
\usepackage{ae,aecompl}

\usepackage{newtxtext,newtxmath}

\title{Connecting Young Star Clusters to CO Molecular Gas in NGC~7793 with ALMA--LEGUS}

\author[K. Grasha et al.]{
K. Grasha,$^{1,2,3}$\thanks{E-mail: kathryn.grasha@anu.edu.au}
D. Calzetti,$^{1}$
L. Bittle,$^{4}$
K.E. Johnson,$^{4}$
J. Donovan Meyer,$^{5}$
\newauthor
R.C. Kennicutt,$^{6}$
B.G. Elmegreen,$^{7}$
A. Adamo,$^{8}$
M.R. Krumholz,$^{2,3}$
M. Fumagalli,$^{9}$
\newauthor
E.K. Grebel,$^{10}$
D.A. Gouliermis,$^{11,12}$
D.O. Cook,$^{13}$
J.S. Gallagher III,$^{14}$
A. Aloisi,$^{15}$
\newauthor
D.A. Dale,$^{16}$
S. Linden,$^{2}$
E. Sacchi,$^{15}$
D.A. Thilker,$^{17}$
R.A.M. Walterbos,$^{18}$
M. Messa,$^{7}$
\newauthor
A. Wofford,$^{19}$
L.J. Smith$^{20}$
\\
$^{1}$Astronomy Department, University of Massachusetts, Amherst, MA 01003, USA; kathryn.grasha@anu.edu.au\\
$^{2}$Research School of Astronomy and Astrophysics, Australian National University, Canberra, ACT 2611, Australia\\
$^{3}$ARC Centre of Excellence for All Sky Astrophysics in 3 Dimensions (ASTRO 3D), Australia\\
$^{4}$Department of Astronomy, University of Virginia, 3530 McCormick Road, Charlottesville, VA 22904, USA\\
$^{5}$National Radio Astronomy Observatory, Charlottesville, VA 22901, USA\\
$^{6}$Institute of Astronomy, University of Cambridge, Cambridge, UK\\
$^{7}$IBM Research Division, T.J. Watson Research Center, Yorktown Hts., NY, USA\\
$^{8}$Department of Astronomy, The Oskar Klein Centre, Stockholm University, Stockholm, Sweden\\
$^{9}$Institute for Computational Cosmology and Centre for Extragalactic Astronomy, Durham University, Durham, UK\\
$^{10}$Astronomisches Rechen-Institut, Zentrum f\"ur Astronomie der Universit\"at Heidelberg, M\"onchhofstr.\ 12--14, 69120 Heidelberg, Germany\\
$^{11}$Zentrum f\"ur Astronomie der Universit\"at Heidelberg, Institut f\"ur Theoretische Astrophysik, Albert-Ueberle-Str.\,2, 69120 Heidelberg, Germany\\
$^{12}$Max Planck Institute for Astronomy, K\"{o}nigstuhl\,17, 69117 Heidelberg, Germany\\
$^{13}$California Institute of Technology, Pasadena, CA, USA\\
$^{14}$Dept. of Astronomy, University of Wisconsin--Madison, Madison, WI, USA\\
$^{15}$Space Telescope Science Institute, Baltimore, MD, USA\\
$^{16}$Department of Physics and Astronomy, University of Wyoming, Laramie, WY, USA\\
$^{17}$Department of Physics and Astronomy, The Johns Hopkins University, Baltimore, MD, USA\\
$^{18}$Dept. of Astronomy, New Mexico State University, Las Cruces, NM, USA\\
$^{19}$Instituto de Astronom\'{i}a, Universidad Nacional Aut\'{o}noma de M\'{e}xico, Unidad Acad\'{e}mica en Ensenada, Km 103 Carr. Tijuana-Ensenada, Ensenada 22860, M\'{e}xico\\
$^{20}$European Space Agency/Space Telescope Science Institute, Baltimore, MD, USA\\
}

\pubyear{2018}

\begin{document}
\label{firstpage}
\pagerange{\pageref{firstpage}--\pageref{lastpage}}
\maketitle

\begin{abstract}
We present an investigation of the relationship between giant molecular cloud (GMC) properties and the associated stellar clusters in the nearby flocculent galaxy NGC~7793. We combine the star cluster catalog from the HST LEGUS (Legacy ExtraGalactic UV Survey) program with the 15 parsec resolution ALMA CO(2--1) observations. We find a strong spatial correlation between young star clusters and GMCs such that all clusters still associated with a GMC are younger than 11~Myr and display a median age of 2~Myr. The age distribution increases gradually as the cluster--GMC distance increases, with star clusters that are spatially unassociated with molecular gas exhibiting a median age of 7~Myr. Thus, star clusters are able to emerge from their natal clouds long before the timescale required for clouds to disperse. To investigate if the hierarchy observed in the stellar components is inherited from the GMCs, we quantify the amount of clustering in the spatial distributions of the components and find that the star clusters have a fractal dimension slope of $-0.35 \pm 0.03$, significantly more clustered than the molecular cloud hierarchy with slope of $-0.18 \pm 0.04$ over the range 40--800~pc. We find, however, that the spatial clustering becomes comparable in strength for GMCs and star clusters with slopes of $-0.44\pm0.03$ and $-0.45\pm0.06$ respectively, when we compare massive ($>$10$^5$~\Msol) GMCs to massive and young star clusters. This shows that massive star clusters trace the same hierarchy as their parent GMCs, under the assumption that the star formation efficiency is a few percent.
\end{abstract}




\begingroup
\let\clearpage\relax
\endgroup
\newpage
 
\section{Introduction}\label{sec:intro}
Star formation is a hierarchical, scale-free process in both space and time \citep{efremov98} spanning from individual stars to entire star-forming galaxies, and is a consequence of giant molecular clouds (GMCs) converting their molecular mass into newly formed stars via fragmentation. The observed hierarchical distribution of stars, star clusters \citep[e.g.,][]{gomez93, zhang01, odekon08, gouliermis14, sun17a, sun18}, and progenitor clouds/cores \citep[e.g,][]{elmegreenfalgarone96, johnstone00, johnstone01, sanchez10}, is believed to be imposed by the hierarchical nature of turbulence \citep{elmegreen96, elmegreen97, padoan02, hopkins13a, hopkins13b} throughout the interstellar medium (ISM). 

The Legacy ExtraGalactic Ultraviolet Survey\footnote{https://archive.stsci.edu/prepds/legus/} \citep[LEGUS, HST GO--13364;][]{calzetti15a}, a Hubble Space Telescope (HST) Treasury program of 50 local ($<$18~Mpc) galaxies observed in the UV and optical regimes have enabled unprecedented investigations into the star formation hierarchies across a diverse population of galaxies, improving our understanding of the processes of star formation in the local universe \citep{elmegreen14, gouliermis15, gouliermis17, grasha15, grasha17a, grasha17b}. 

The dynamical structure of a galaxy is capable of impacting the local star formation process \citep[e.g.,][]{renaud13}, and consequently, the overall organization and survival of star-forming structures. As such, high-resolution studies of local galaxies are ideal to study the complex interplay between gas and the subsequent impact on star formation. Early results from the Plateau de Bure Interferometer Arcsecond Whirlpool Survey \citep[PAWS;][]{schinnerer13, pety13} reveal that the spiral structure of M~51 impacts the gas density of the local environment, which alters the properties of the GMCs. This inevitably impacts the overall organization and structure of the ISM at scales of individual GMCs \citep{hughes13} and the resulting products of star formation \citep{messa18b}. 

This paper focuses on the flocculent galaxy NGC~7793, one of the closest galaxies in the LEGUS sample. A primary science goal of LEGUS is to further our understanding of the connection between localized sites of star formation and the global star formation process. LEGUS has vastly increased the number of high-quality catalogs of young star clusters over a broad range of galactic environments, improving our understanding of the parameters responsible for star cluster formation and evolution in a homogeneous and consistent manner (Kim \etal\ in prep). The lifetimes of the star-forming complexes were previously investigated for NGC~7793 in \citet{grasha17a, grasha17b}, where we find relatively small size scales of $\sim$200 parsec for the correlation length of star formation. 

The goal of this paper is to investigate the connection between molecular gas and associated stellar populations, building upon the seminal work of \citet{leisawitz89} where they studied the effect of stars on the molecular clouds within the Milky Way. We previously examined the star cluster and GMC association in another LEGUS galaxy, the interacting spiral system NGC~5194 \citep{grasha18}. This paper complements the NGC~5194 study with recent CO (J = 2--1) observations of NGC~7793 (Bittle \etal\ in prep) from the Atacama Large Millimeter/submillimeter Array (ALMA), serving as an excellent comparison to connect the properties of star clusters and molecular gas to the spiral structure in two different galactic systems. 

NGC~7793 has 25'' resolution CO(J = 3--2) observations with the Atacama Submillimeter Telescope Experiment\citep{muraoka16}; the 0\farcs85 resolution provided by ALMA allow us to resolve down to sub-GMC scales and examine the molecular gas at scales comparable to the star clusters. NGC~7793 is a cluster-poor flocculent galaxy, which makes it an interesting system to study the properties of star clusters in the context of their connection to the spiral structure and the timescale for star formation to remain associated with molecular clouds. These results are quite relevant to the longstanding issue of molecular cloud lifetimes, both before and during star formation, and its relation to the environment \citep[e.g.,][]{kawamura09, miura12, whitmore14, heyer15, kreckel18}.

The paper is organized as follows. The galaxy selection and reduction process and the acquisition and reduction of the ALMA data are described in Section~\ref{sec:sample}. The cluster selection and identification process is described in Section~\ref{sec:clusterselection}. The correlation of the star clusters to the molecular gas is described in Section~\ref{sec:results} and we quantify the hierarchy of the star clusters and the GMCs in Section~\ref{sec:2pcf}. Finally, we summarize the findings of this study in Section~\ref{sec:summary}.

\section{NGC 7793}\label{sec:sample}

\subsection{The HST UV/Optical Observations}\label{sec:7793}
\begin{figure*}
\includegraphics[width=0.99\textwidth]{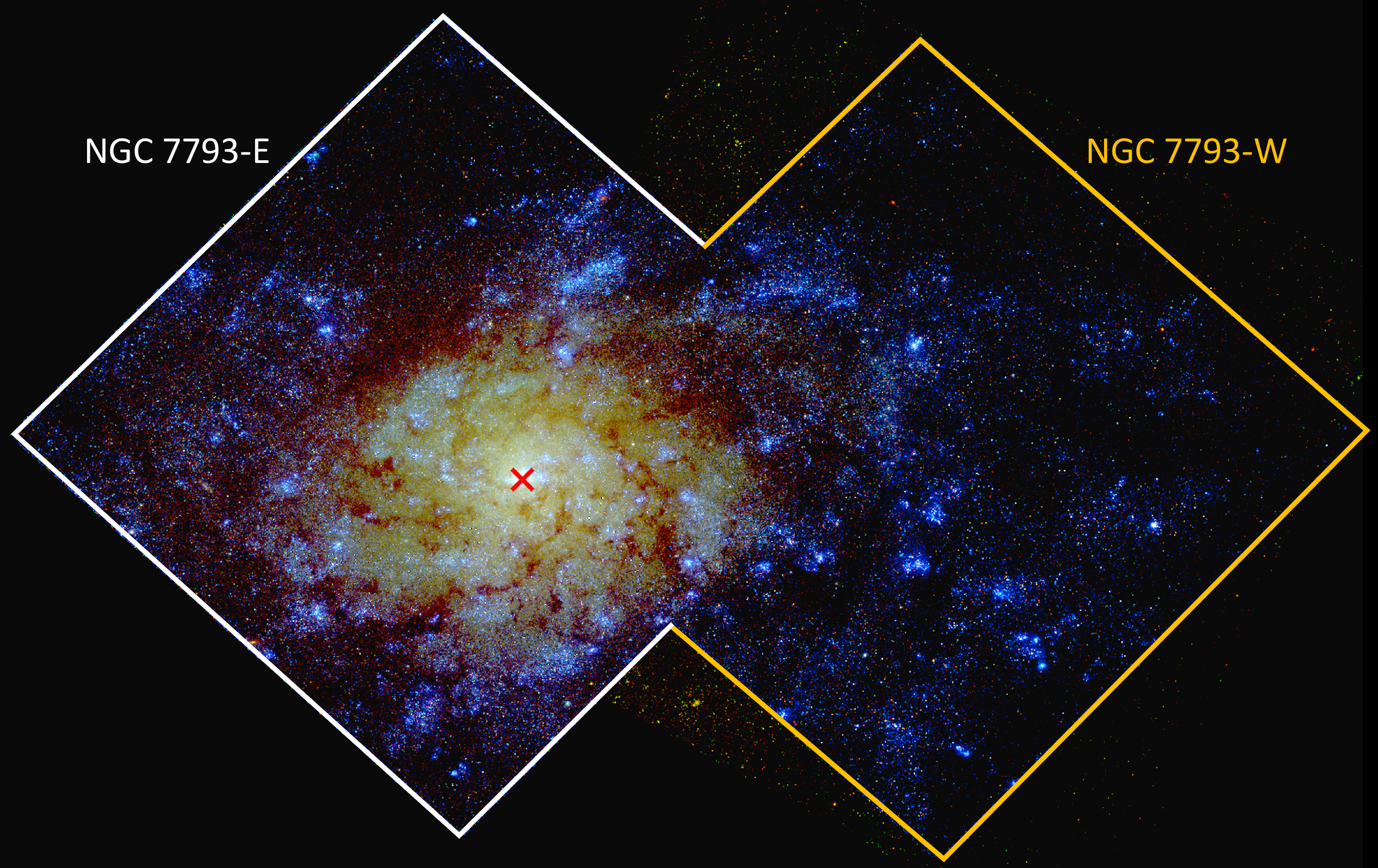} 
\caption{
LEGUS color composite mosaic of NGC~7793: UVIS/F275W and UVIS/F336W (blue), UVIS/F435W and UVIS+ACS/F555W (green), and UVIS+ACS/F814W bands (red). The outlines show the HST/WFC3 field of view, with the east pointing (NGC~7793-E) shown in white and the west pointing (NGC~7793-W) shown in yellow. The red cross marks the center of the galaxy.
} 
\label{fig:rgb}
\end{figure*}

In this paper, we select NGC~7793 from the LEGUS survey and observe the molecular gas with ALMA (Section~\ref{sec:alma}) to study the impact of spiral arm structures on the properties of both star clusters and the molecular gas reservoir. NGC~7793 is a nearby \citep[$3.44\pm0.15$ Mpc assuming a distance modulus of $27.68\pm0.09$;][]{pietrzynski10} SAd flocculent galaxy in the Sculptor group. It is characterized by diffuse, broken spiral arms, has no bar, and a very faint central bulge (Figure \ref{fig:rgb}). It has a relatively low star formation rate \citep[SFR(UV) $\sim$0.52~\Msol~yr$^{-1}$;][]{calzetti15a}, a central oxygen abundance of 12 + log (O/H) = $8.50\pm0.02$ and an oxygen abundance gradient of $-$0.0662 $\pm$ $-$0.0104 dex/kpc \citep{pilyugin14}. \citet{kahre18} finds a small dust--to--gas ratio that does not change with galactocentric distance. NGC~7793 has a stellar mass of $3.2\times10^9$~\Msol\ and is inclined 47 degrees \citep{carignan90}, has an isophotal R$_{25}$ radius of 4.65~kpc and a maximum rotational velocity of 116~\kms\ \citep{dicaire08}. 
%

The HST observations of NGC~7793 cover two pointings, the western and eastern part of the galaxy. The eastern pointing is observed in all five bands (F275W, F336W, F438W, F555W, F814W) with the WFC3. The west pointing is observed in only three bands (F275W, F336W, F438W) with the remaining two bands (F555W, F814W) taken from archival ACS observations (GO--9774; P.I. S.S. Larsen). All archival ACS images are re-reduced using the same pipeline as the UV and U images with WFC3/UVIS from the LEGUS project. Reduced science frames are drizzled to a common scale resolution, to match the native WFC3 pixel size (0.0396 arcsec/pixel), corresponding to a resolution element of 0.66 parsec/pixel. The frames have all been aligned and rotated with North up. We refer the reader to the detailed descriptions of the standard reduction of the LEGUS data sets in \citet{calzetti15a}. 

\begin{figure*}
\includegraphics[width=\textwidth]{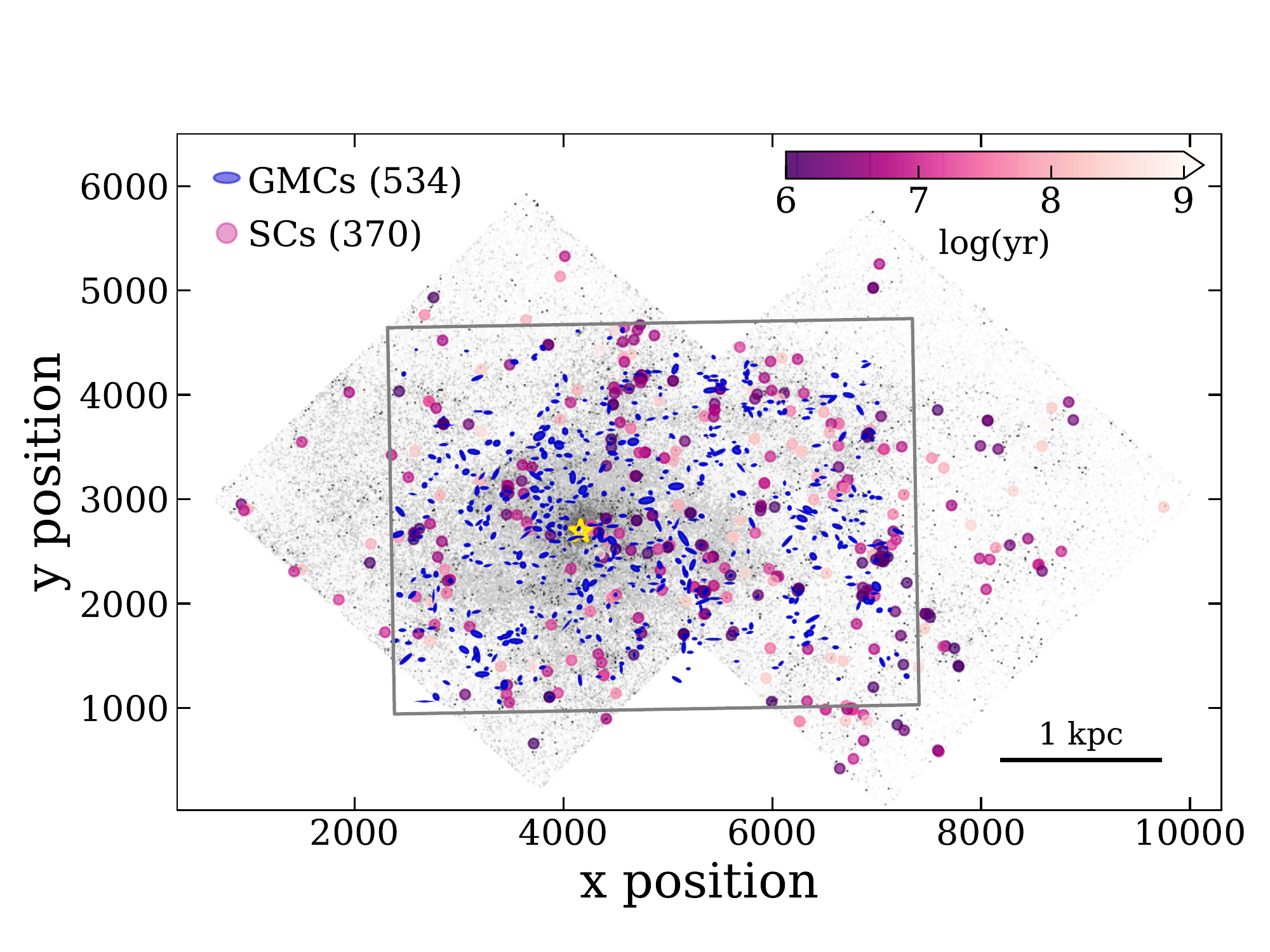} 
\caption{
Positional location and sizes of the GMCs (blue ellipses, minimum mass 3650~\Msol, maximum mass $8.4\times10^5$~\Msol) and star clusters (circles) in NGC~7793 over the UVIS/F438W image. The star clusters are colored according to their ages, where the youngest are darker pink and the oldest are white. The yellow star shows the center of the galaxy. The gray rectangle shows the outline of the ALMA coverage. Star clusters not located within the ALMA coverage are excluded from all star cluster--GMC comparisons.} 
\label{fig:7793}
\end{figure*}

Figure~\ref{fig:7793} shows the gray-scale V-band image of both the east and west UVIS pointings overlaid with the positions of the GMCs and ALMA coverage (Section~\ref{sec:alma}) and the star clusters (Section~\ref{sec:clusterselection}). 

\subsection{The ALMA CO Observations}\label{sec:alma}
A detailed quantification of the data acquisition, reduction, and creation of the GMC catalog, along with the properties of the molecular clouds (mass function, power spectrum, etc.) and how they depend on their location within NGC~7793, are described in Bittle \etal(in prep). We list here the details necessary for this study.

We observed the CO (2--1) transition with ALMA band 6 (230.36366 GHz) in the inner 180''$\times$114'' ($3\times2$~kpc) of NGC~7793 (ALMA programs 2015.1.00782 \& 2016.1.00674; PI: K.E. Johnson). The total integration time is 3 hours with the 12-m array with 149 pointings at 72.6 seconds of integration per pointing with a mosaic spacing of 12\farcs9 between pointings. Our angular resolution of 0\farcs85 allows us to resolve sub-GMC sizes (8~pc scales) and we are sensitive to emission on scales as large as 11'' ($\sim$175~pc), which allow us to recover the largest molecular complexes in the galaxy. We have sufficient sensitivity to detect molecular clouds with masses of $\sim$10$^4$~\Msol. We also achieve a velocity resolution of 1.2~\kms, sufficient to resolve individual clouds with expected velocity dispersions of order 2--3~\kms. The final rms noise of the observations is $\sim$4 mJy/beam. The non-detection of spatially extended emission should not affect any of the results presented in this paper.

\subsection{Creating the GMC Catalog}\label{sec:gmc}
We create the GMC catalog from the CO(2-1) position-position-velocity data with the CPROPS segmentation algorithm for spectral line emission \citep{rosolowsky06}. After identifying a typical cube rms noise, local maxima are identified above 4-$\sigma$. Emission above 2-$\sigma$ that lie in at least two continuous velocity channels around an identified peak is then assigned to that peak. For a local maximum within a kernel range of another local maximum, we define a shared contour level. If the peaks both lie above 3.5-$\sigma$ of the isophote, they are determined to be unique entities and the emission is separated appropriately. From these finalized identifications of clouds, we are able to measure properties such as size ($R_{GMC}$), line width ($\sigma_v$), and luminosity ($L_{CO}$) for each identified cloud. 

\begin{figure*}
\includegraphics[width=\textwidth]{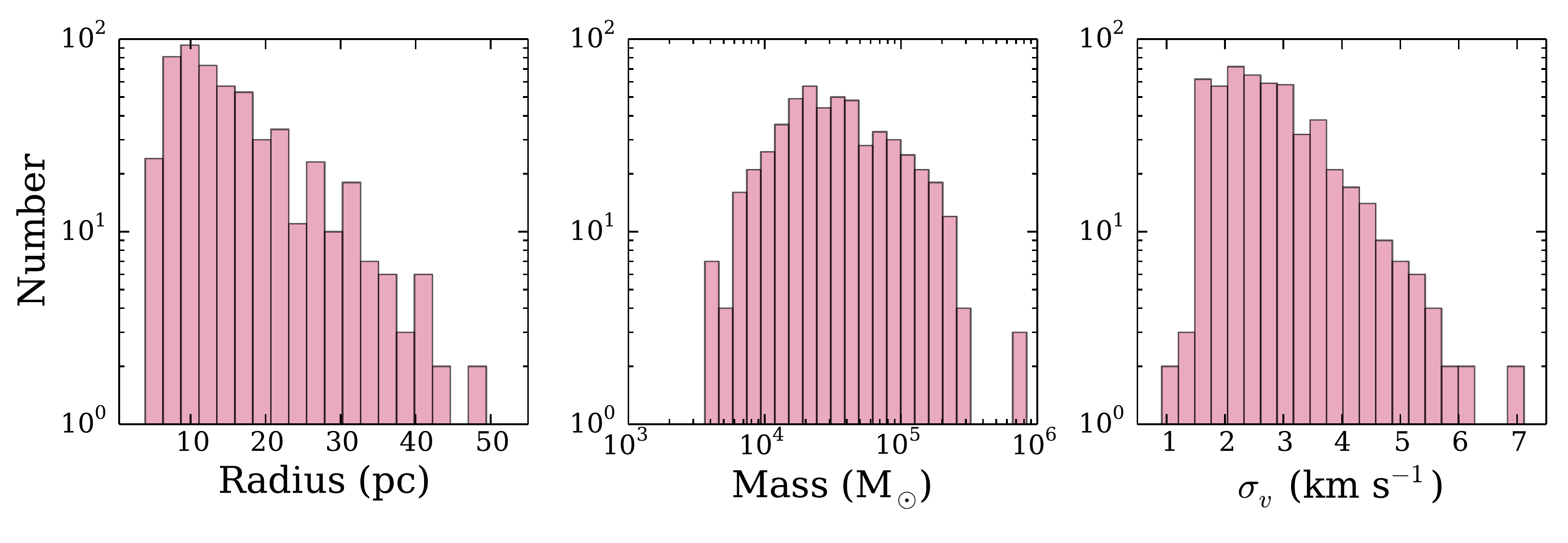}
\caption{
Distribution of the radius, mass, and velocity dispersion of the clouds in NGC~7793.  \label{fig:gmcprops}}
\end{figure*}

By construction, the GMCs represent significant peaks in the CO emission and we assume that these correspond to the cluster-forming structures. Using the position angle, radius, and semimajor axis, we represent the GMCs as ellipses in Figure~\ref{fig:7793}. The mean radius of the GMCs in this study is 16~pc. Figure~\ref{fig:gmcprops} shows the distribution of the mass, radius, and velocity dispersion for the clouds. The final GMC catalog for NGC~7793 will be presented in Bittle \etal\ (in prep).

\section{Creating the Visually-Identified Star Cluster Catalogs}\label{sec:clusterselection}
A detailed description of the cluster selection, identification, photometry, spectral energy distribution (SED) fitting procedures, and completion limits for the LEGUS galaxies is presented in \citet{adamo17}. We summarize here briefly the aspects that are important for the current paper.

\subsection{Star Cluster Selection}\label{sec:SCselection}
The LEGUS process of producing cluster catalogs is a multi-step process, with an initial automated extraction of the cluster candidates (Section~\ref{sec:SCselection}), followed by a visual identification of a sub-sample of the brightest clusters to remove contaminants (Section~\ref{sec:SCidentification}) in order to create the final science-ready, visually identified cluster catalog. 

The initial {\it automated catalog} of star cluster candidates is extracted from a white-light image generated by using the available photometric bands \citep[see][]{calzetti15a} with source extractor \citep[SExtractor;][]{bertin96}. The SExtractor parameters are optimized to extract sources with at least a 3$\sigma$ detection in a minimum of 5 contiguous pixels. This automatic procedure returns the positions of candidate clusters within the image and the concentration index (CI; magnitude difference of each source within an aperture of 1 pixel compared to 3 pixels). The CI is related to the size of stellar systems \citep{ryon17} and can be used to differentiate between individual stars and stellar clusters. Star clusters, in general, have more extended light profiles, and therefore, larger CI values compared to individual stars. The CI reference value used to distinguish between unresolved sources (stars) and resolved sources (candidate clusters) within NGC~7793 is 1.3 mag; we disregard any sources with a CI value less than this reference value. The typical size distribution of star clusters peaks at $\sim$3~pc \citep{ryon17}; as demonstrated in \citet{adamo17}, our cluster detections are complete down to 1~pc in size out to 10~Mpc, well below the peak of their size distribution. Thus, the cluster population in NGC~7793, which at a distance of only 3.44~Mpc, is expected to be complete well below the size distribution peak. 

Standard photometry is performed on the cluster candidates using a science aperture radius of 5 pixels and a sky annulus at 7 pixels with a width of 1 pixel. Average aperture corrections are estimated using a cluster control sample \citep[see][]{adamo17}, estimated as the difference between the magnitude of the source within a 20 pixel radius with a 1 pixel sky annulus minus the magnitude of the source at the science aperture. These average aperture corrections in each filter are applied to the standard photometry of all the clusters. Corrections for foreground Galactic extinction \citep{schlafly11} are also applied to the photometry. All positions are corrected for an inclination of 47.4 degrees. 

All cluster candidates detected in at least four of the five bands with photometric error $\leq$0.3~mag undergo spectral energy distribution (SED) fitting procedures and error propagation as described in \citet{adamo10, adamo12} to extract the age, mass, and color excess \ebv\ of each source. The SED fitting analysis is performed with Yggdrasil single stellar population (SSP) models \citep{zackrisson11}. The Yggdrasil spectral synthesis code combines the deterministic stellar population synthesis models of Starburst99 \citep{leitherer99} with the photoionized nebular emission predicted by Cloudy \citep{ferland98, ferland13}. All cluster catalogs for the LEGUS galaxies implement a \citet{kroupa01} universal initial mass function (IMF). Our SED fits assume that the IMF is universal, i.e., there are no variations that depend on the mass, age, or other characteristics of the stellar clusters. This is in agreement with the results of \citet{calzetti10, andrews13} and \citet{andrews14}, where the authors find that star cluster populations in galaxies are consistent with a universal stellar IMF.  Conversely, \citet{kirk12, marks12, ramirez16} and \citet{stephens17}, indicate that there may be systematic variations in the stellar IMF along the models of \citet{weidner13}. Our data do not enable us to discriminate among these possibilities, as discussed in \citet{ashworth17}, and this remains a potential difficulty in our analysis. However, as described below, tests using stochastic IMF sampling do not yield significantly different results from a deterministic, universal IMF, for the star cluster population in NGC 7793. For NGC~7793, the cluster catalog SED is derived using the Padova-AGB SSP stellar isochrone tracks \citep{vazquez05} and a starburst attenuation curve \citep{calzetti00} with the assumption that stars and gas undergo the same amount of reddening. For wavelengths longer than $\sim$3000~\AA, attenuation and extinction dust models are similar in shape, and thus, the adopted dust model does not greatly impact the results. 

The star cluster properties of NGC~7793 are also derived using a Bayesian analysis method with stochastically sampled cluster evolutionary models \citep{krumholz15a} using the Stochastically Lighting Up Galaxies \citep[SLUG;][]{dasilva12, krumholz15b} code. SLUG returns the full posterior probability distribution function (PDF) of the physical properties of each cluster rather than a single best fit. This approach does not assume a fully sampled IMF, which becomes especially relevant for the accurate derivation of cluster properties in cluster masses below $\sim$ $10^{3.5}$~\Msol\ \citep{cervino02}, which is near the completeness limit of the LEGUS catalogs for cluster ages up to 200~Myr at distances of $\sim$10~Mpc \citep{adamo17}. NGC~7793 is located at $3.44\pm0.15$~Mpc, and as a result, we expect the cluster catalog to be complete to significantly less massive clusters. We examine the effects that the method of deriving the ages with deterministic, universal-IMF models has on our results and find that the science results appear to be relatively insensitive whether the properties are derived deterministically or stochastically. As a result, in this paper we only consider the cluster properties that are derived with deterministic models that assume a universal IMF. We refer the reader to \cite{adamo17} and references therein for the full description of the deterministic models use to derive the LEGUS cluster catalogs that we briefly described above.

\subsection{Visual Inspection and Star Cluster Classification}\label{sec:SCidentification}
After the extraction of the clusters and the SED fitting procedure, all clusters with an absolute magnitude brighter than $-6$ mag in the V-band undergo visual inspection by a minimum of three independent classifiers within the LEGUS team to secure the final {\it visual catalog}. This $-6$ magnitude limit is defined by the detection limits of the LEGUS sample and enables selection down to a $\sim$1000~\Msol, 6~Myr old cluster with \ebv\ = 0.25 \citep{calzetti15a}. 484 cluster candidates are brighter than the magnitude cut off and undergo the visual classification procedure. 

The visual classification is performed on a V-band and three-color composite image in addition to the surface contours, radial profiles, and surface plots of each source \citep[see][]{adamo17}. The LEGUS cluster classification is based on the morphology and color of each source and is necessary in order to exclude non-cluster contaminants within the automatically extracted catalog. Each sources gets classified under one of four classes: (1) symmetric and centrally compact star clusters, usually uniform in color; (2) compact, asymmetric star clusters with some degree of elongation, usually uniform in color; (3) multiple-peaked profiles that show an underlying diffuse emission, color gradients are common; and (4) non-clusters, including but not limited to foreground stars, asterisms, background galaxies, saturated/bad pixels, etc. The final cluster catalog for NGC~7793 contains 370 clusters of class 1, 2, and 3. 

In general, class 1 and 2 clusters are older and are potentially gravitationally bound systems with ages greater than their crossing time \citep{ryon17} whereas class 3 objects are in general much younger, and due to their multi-peak nature, we refer to these systems as compact associations. Prior work in other LEGUS galaxies show that the morphological classification may contain information about the dynamical state of the clusters as well \citep{adamo17, grasha17a}. The class 3 associations have spatial distributions that are significantly more clustered compared to the distribution of class 1 and 2 clusters. In this work, we compare the total star cluster sample to that of the GMCs. NGC~7793 is one of the closest galaxies in the LEGUS sample, and as such, we have sufficient resolution to resolve down to very compact or loose associations that may otherwise be missed in more distant systems. 42\% of the clusters in the NGC~7793 catalog are class 3 associations, among the highest rate out of all the LEGUS cluster catalogs \citep[see Figure 2 of][]{grasha17a}. Thus we are confident that we are efficient at identifying what may be considered young and loose stellar aggregates within NGC~7793.

The final star cluster catalog for NGC~7793, as well as the rest of the LEGUS star cluster catalogs, is available online\footnote{https://archive.stsci.edu/prepds/legus/dataproducts-public.html} on the Mikulski Archive for Space Telescopes (MAST).

\section{Results and Analysis}\label{sec:results}

\subsection{Separation of Young Star Clusters and GMCs}\label{sec:gas}
To connect the young star clusters to their environment, we compare the projected spatial locations of the clusters to the projected location and sizes of peaks in the molecular gas. Previous observations show that stars and star clusters quickly become unassociated with the GMCs from where they are born, either due to drift or from gas expulsion. Within the Antennae galaxy, star clusters are exposed at $\sim$5~Myr \citep{whitmore14, matthews18} as well as within M~33, where exposed clusters show a peak around 5~Myr with an embedded phase that lasts only 4~Myr \citep{corbelli17}. Within Milky Way molecular clouds, the lifetimes of various star formation phases can be exceptionally short, $\la$1~Myr \citep{battersby17}. Determining the timescales for the association of GMCs with star formation and how quickly star clusters separate them from their natal origins and the dependency, if any, on the galactic environment thus provide a more complete picture of star formation. Within this paper, if a star cluster and a GMC overlap in projection, we consider that pair an association.

\subsubsection{Shortest Distance between Clusters and GMCs}
\begin{figure}
\includegraphics[scale=0.42]{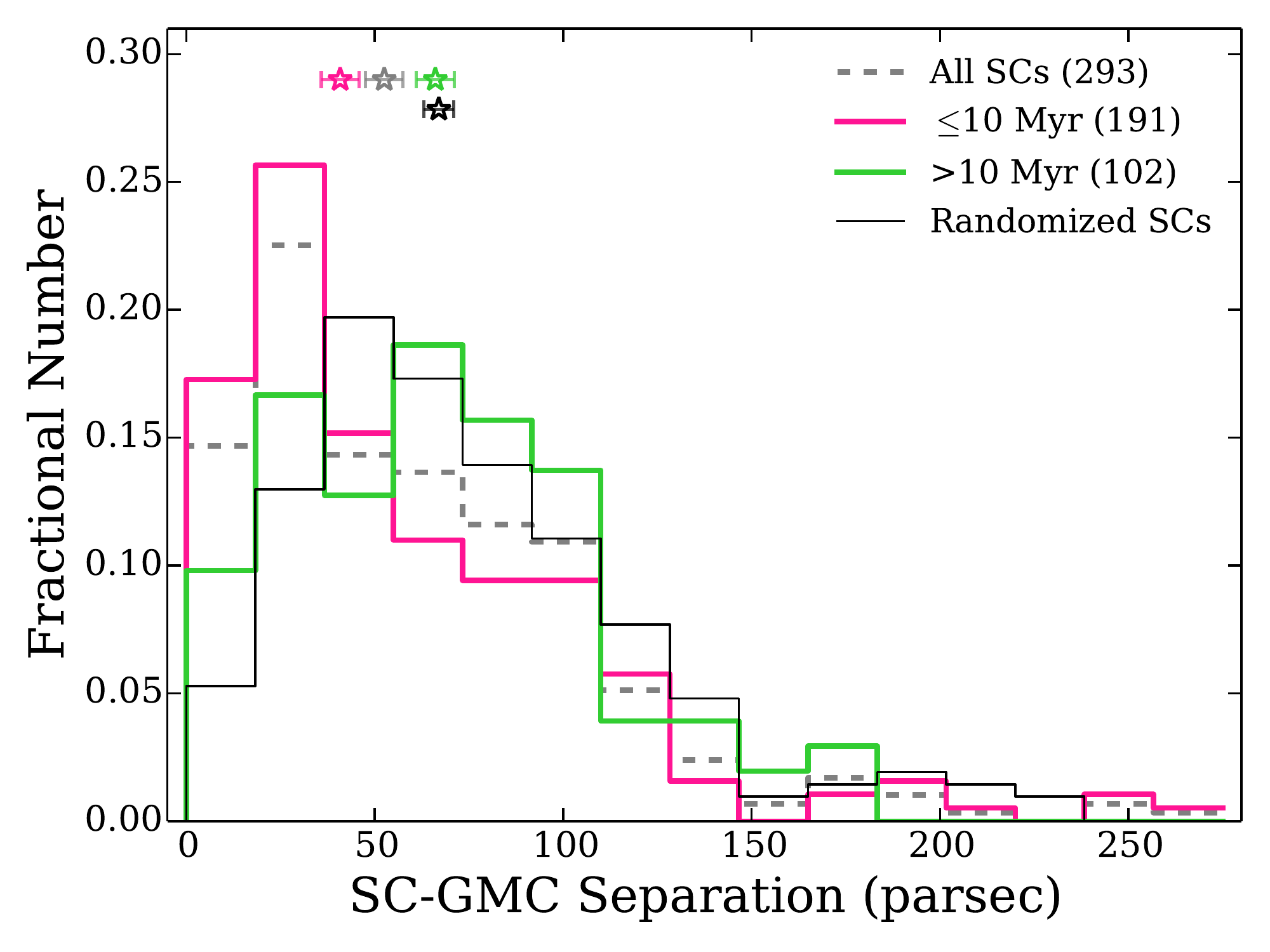}
\caption{
Fractional distribution of the shortest distance for each star cluster to the closest GMC divided into two age bins. The star symbols show the median value for each distribution along with the 1$\sigma$ uncertainties from bootstrap estimates based on 10,000 samples. The global average (dashed gray) for the shortest distance between each star cluster to the closest GMC is $53\pm5$ parsec. Star clusters with ages less than 10~Myr (pink) show shorter distances of $41\pm4$~pc compared to ages older than 10~Myr (green) at $66\pm5$~pc. Younger star clusters are spatially closer to a GMC than older star clusters and the difference is significant given the scatter. There is a clear excess observed for the youngest ($<$10~Myr) star clusters based on the expected distribution if the positions of the clusters are randomized, while preserving their radial density distribution (thin black line, median value of $67\pm4$~pc). 
\label{fig:sto}}
\end{figure}

We take all the star clusters within the ALMA coverage, reducing the total clusters in the catalog from 370 to 293, and measure the shortest distance to the center of the nearest GMC. Figure~\ref{fig:sto} shows the distribution of the shortest star cluster -- GMC distance. For the entire sample, the median of the distances is $53\pm5$~pc. The youngest star clusters (ages less than 10~Myr) have a median star cluster--GMC distance that drops to $41\pm4$~pc and the remaining star clusters that are older than 10~Myr have a median cluster--GMC pair distance of $66\pm5$~pc. Younger star clusters are significantly closer in proximity to a GMC than older star clusters despite the small numbers of star clusters and GMCs in NGC~7793. 

In Figure~\ref{fig:sto} we also show the expected distribution expected if the positions of the star clusters are randomized, while preserving their radial density distribution, as the spatial density of the star clusters is higher at smaller galactocentric distances. For the observed clusters, there is a clear excess of the clusters with ages $\la$10~Myr in proximity of the GMCs. The older star clusters ($>$10~Myr) show much weaker correlations, with distributions that are almost consistent what is expected from a random cluster population. The same weak spatial correlation between star clusters and GMCs for ages older than 10~Myr is seen within the LMC \citep{kawamura09}.

\begin{figure}
\includegraphics[scale=0.42]{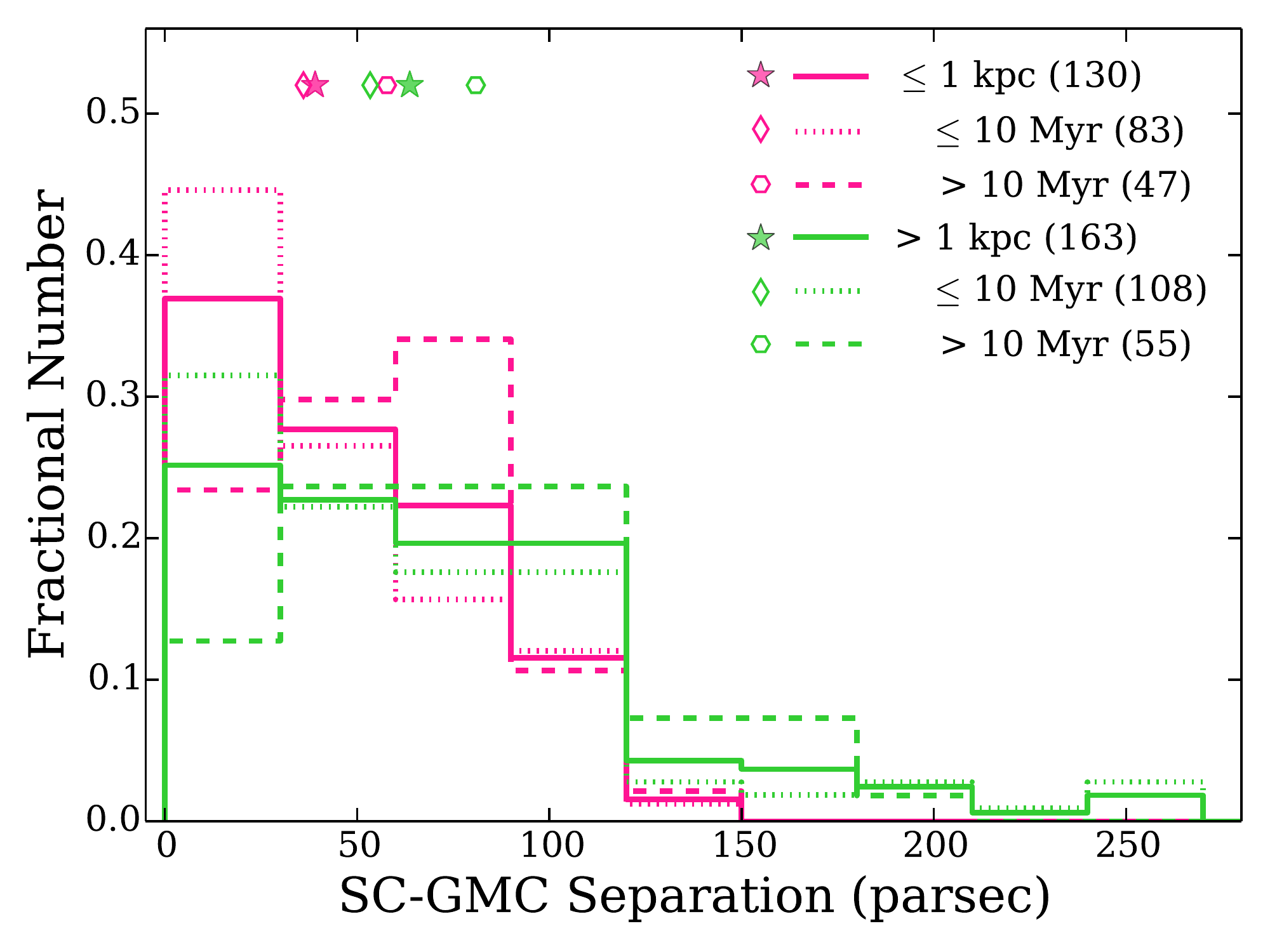}
\caption{
Fractional distribution of the shortest distance for each star cluster to the closest GMC broken into galactocentric radial bins and age bins. The symbols show the median value for each distribution. Star clusters less than a galactocentric radius of 1~kpc (pink) are on average $39\pm4$~pc from the nearest GMC with ages less than 10~Myr (dotted pink line) showing slightly shorter mean distances of $36\pm4$~pc compared to ages greater than 10~Myr (dashed pink line) at $58\pm10$~pc. Star clusters at galactocentric distances greater than 1~kpc (green) are on average $64\pm7$~pc from the nearest GMC, with mean distance decreasing for the youngest ($\leq$10~Myr; dotted green line) to $53\pm9$~pc and increasing to $81\pm8$~pc for the older clusters ($>$10~Myr; dashed green line). On average, younger star clusters are spatially closer to a GMC than older star clusters although the distance from the center of the galaxy has a larger impact with the average SC--GMC distance increasing significantly with increasing galactocentric distance. Errors are bootstrap estimates from 10,000 samples and are not shown on the plot.
\label{fig:sc_gmc_mindistance}}
\end{figure}

The galactocentric distance $r_{GC}$ of the closest star cluster/GMC pairs has a bigger influence on the age difference and we investigate this effect by dividing the sample in half at 1~kpc. Clusters located closer than $r_{GC}<1$~kpc have a median SC--GMC separation of only $39\pm4$~pc, rising to $64\pm7$~pc for clusters beyond 1~kpc (Figure~\ref{fig:sc_gmc_mindistance}). The density of star clusters is higher in the central kpc of the galaxy, thus separations are statistically smaller in the inner 1 kpc region of the galaxy; this is the primary driver of the trend for increasing separation between SC--GMC pairs with increasing galactocentric distance. The excess for SC--GMC separations less than 50 pc at galactocentric distances less than 1 kpc can be best seen by comparing the solid pink line (<1 kpc distances) and solid green line (>1 kpc distances) in Figure~\ref{fig:sc_gmc_mindistance}. 
%
%

\subsection{Properties of Star Clusters Associated with GMCs}\label{sec:SCs_insideGMCs}
With the proper multi-wavelength observational data set, we can quantify the timescale of emergence, and thus the velocity required for star clusters to become unassociated with their natal molecular clouds. We investigate the disassociation timescale by tracking how the distribution in the age of the cluster populations changes as a function of their distance from the center of every GMC. We do not allow double counting and each cluster can be assigned to only one GMC. In situations where multiple clouds lie on top of each other, the star cluster is assigned to the closest GMC. For all but one star cluster, the closest GMC is also the most massive cloud. 

Within NGC~7793, we find 13 star clusters (4\%)  lie within the extent of 12 GMCs (2\%). 31 clusters are located at distances just outside their nearest GMC ($1\times R_{\rm GMC}$), but less than 2 radii of a GMC center. 26 star clusters located at distances greater than 2 but less than 3 radii away from their nearest GMC. The remaining 224 star clusters (76\%) are unassociated with any cloud (i.e., located at distances greater than $3\times R_{\rm GMC}$ from the center of the nearest GMC). We select $3\times R_{\rm GMC}$ as the definition of an ``unassociated'' that is the distance where the star clusters show median ages that are above that of the global star cluster population. 

\begin{figure}
\includegraphics[scale=0.48]{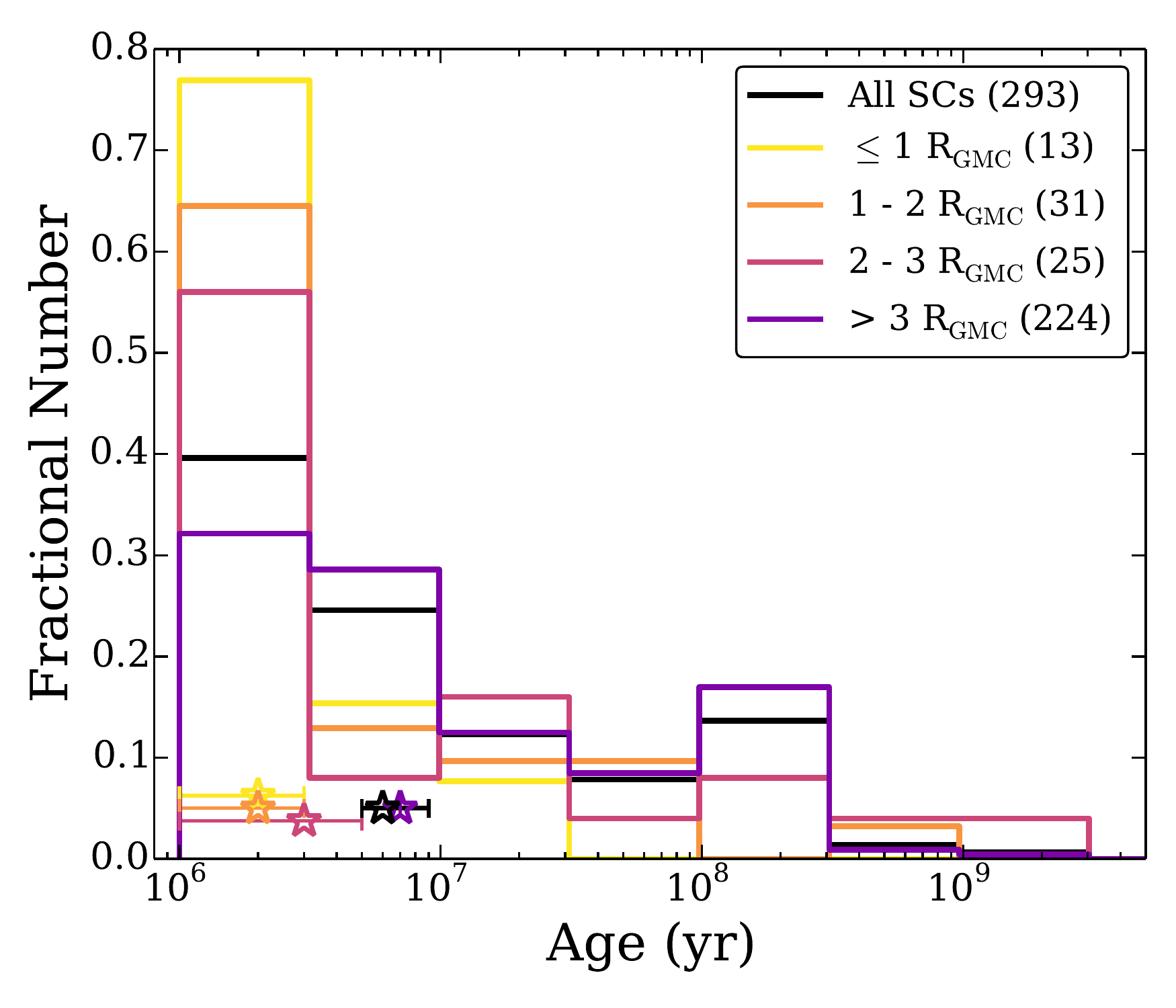}
\caption{
Distribution of the star cluster ages for the entire sample (black), star clusters located $\leq$1 $R_{\rm GMC}$ (yellow), within 1--2 $R_{\rm GMC}$ (orange), 2--3 $R_{\rm GMC}$ (pink), and star clusters unassociated with a GMC ($>$3 $R_{\rm GMC}$; purple). The stars show the median age of each distribution ($6\pm1$~Myr, $2\pm1$~Myr, $2\pm1$~Myr, $3\pm2$~Myr, and $7\pm1$, respectively; offset in the y-direction) and the 1$\sigma$ uncertainties are bootstrap estimates based on 10,000 samples. Star clusters located within a GMC are generally much younger than star clusters not within close proximity to a GMC and the age increases with increasing distance from the GMC. 
\label{fig:dist_agemass}}
\end{figure}

Figure~\ref{fig:dist_agemass} shows the distribution of the cluster ages for those with and without associations to GMCs. The median age of all clusters is 6~Myr, which is 3 times older than clusters located in proximity to GMCs at both $<$1~$R_{\rm GMC}$ and 1--2~$R_{\rm GMC}$. Clusters that are between 2 and 3 radii from a GMC have median ages of 3~Myr. Star clusters that are unassociated with any GMC are on average older than all clusters and those associated with GMCs. The age range sampled by the cluster population is small, but the clear excess in the youngest clusters (Figure~\ref{fig:dist_agemass}) at the shortest distances (Figure~\ref{fig:sc_gmc_mindistance}) suggests that the spatial correlation of clouds and young stellar clusters that is lost rapidly. The young ($\la$5~Myr) star clusters in this study can have age uncertainties that are a considerable fraction of the actual assigned value. It is difficult to assess a true difference between 2~Myr and 3~Myr old clusters given our age uncertainties, but the median age changing dramatically from 2--3~Myr clusters closest to a GMC to 7~Myr for those greater than $3R_{\rm GMC}$ is a reliable measurement. We estimate the 1$\sigma$ uncertainties on the median age estimates from bootstrap measurements based on 10,000 samples. There is a relatively low scatter despite the small age range and corroborates the significance in the observed age trend (significant at the 3.4--$\sigma$ level due to very small cluster numbers). There is little or no correlation between the mass of the GMCs and that of the star clusters. We do perform checks by repeating all calculations with a mass cut at 5000~\Msol\ and find that our results are not biased with the inclusion of low-mass clusters. 

Table~\ref{tab:1} lists the star clusters and their properties as a function of distance from their nearest GMC. The trend for younger clusters to be in close proximity to molecular clouds is expected as these exposed stellar systems have started to evacuate their surrounding natal material, but have not lived long enough to either travel far enough to have erased the imprint of their birth location or have completely cleared away the molecular material in their immediate vicinity \citep[e.g.,][]{corbelli17}. The ages we recover here are younger than typical GMC dissolution timescales of $\sim$10--30~Myr \citep{murray11, dobbs13, heyer15}. 

We find that the distribution of \ebv\ values for a given star cluster are unaffected by their proximity to a GMC. The global median for all star clusters has an $A_V$ of 0.49 assuming a starburst attenuation curve. Star clusters that are still associated with a GMC exhibit median $A_V$'s of 0.45 whereas star clusters unassociated with GMCs ($>3 R_{\rm GMC}$) actually show slightly larger values of 0.51. The scatter in \ebv\ for a given age range is considerable and correlates poorly with age \citep{bastian05, adamo10, messa18a}. The difference in age between star clusters inside/outside GMCs therefore cannot be explained by significantly higher extinction affecting the star clusters within the spiral arms or GMCs as the star clusters are already exposed, as shown by the resulting $A_V$ values, indicating the result discussed in the previous paragraph is significant. 

For all 69 star clusters located at distances less than $3 R_{\rm GMC}$, we take the distance of each star cluster from the center of its nearest cloud and divide by the current age of the star cluster. The resulting velocity of 6.2 $\pm$ 0.9~\kms\ is larger than the velocity dispersion in individual GMCs (Figure~\ref{fig:gmcprops}). This velocity primarily describes the speed of the ionization (feedback) that erodes the cloud in an expanding HII region with a component of dynamical motions and cluster drift. The young star clusters will naturally erode the GMCs in which they are embedded, creating growing cavities caused by the winds and ionizing radiation of their massive stars \citep[see, e.g.,][]{lada87}. However, these are effects visible in the vicinity of the clusters while the remainder of the GMC is not affected. 
%
\begin{table*}
 \caption{Properties of star clusters depending on their association with a GMC. Columns list the:
(1) radial distance of the star clusters from a GMC; 
(2) number of star clusters; 
(3) number of GMCs; 
(4) median age of the star clusters; 
(5) median mass of the star clusters; 
(6) median \ebv\ values; 
(7) median GMC radius; and
(8) median GMC mass. The 1$\sigma$ uncertainties are bootstrap estimates based on 10,000 samples. }
 \label{tab:1}
 \begin{tabular}{lccccccc}
  \hline
  Region & N$_{\rm SC}$ & N$_{\rm GMC}$ & Age & Mass$_{\rm SC}$ & \ebv & Radius & Mass$_{\rm GMC}$ \\
   & & & (Myr) & (\Msol) & (mag) & (pc) & (10$^4$ \Msol) \\
  \hline
$\leq$ 1$R_{\rm GMC}$		& 13		& 12		& 2(1) 		& 540(280) 		& 0.45(0.15) & 12.2(1.6) & 4.1(1.0) \\
1-- 2 $R_{\rm GMC}$			& 31		& 22		& 2(1)		&	1100(370) 	& 0.53(0.17) & 20(2) & 9.1(1.6) \\
2--3 $R_{\rm GMC}$			& 25		& 20 	& 3(2)		&	690(490) 	& 0.32(0.15) & 20(4) & 4.2(1.7) \\
Unassociated 						& 224	& 492	& 7(1)		&	1150(260) 	& 0.51(0.04) & 13.2(0.4) & 3.0(0.2) \\
Total 									& 293	& 534	& 6(1)		&	1040(212) 	& 0.49(0.04) & 13.4(0.4) & 3.2(0.2) \\ 
  \hline
 \end{tabular}
\end{table*}

\subsection{Comparison to Previous Work}\label{sec:comparison}
\subsubsection{The Whirlpool Galaxy NGC 5194}\label{sec:5194}
In \citet{grasha18} we performed a similar study to constrain the timescales of the association of GMCs and star clusters in NGC~5194 (7.7~Mpc) with the LEGUS cluster catalog ($\sim$ 1300 star clusters) and the GMC catalog ($\sim$1300 GMCs) at 40~pc resolution \citep{colombo14a} in the inner $6\times9$~kpc of the galaxy with the Plateau de Bure Interferometer Arcsecond Whirlpool Survey \citep[PAWS;][]{schinnerer13,pety13} dataset. 

The median age of the star clusters in NGC~5194 is 30~Myr, five times older than the median population in NGC~7793. The increase in the total number of clusters by 4.7 times allows for a clearer trend and a stronger constraint on the disassociation timescale in a different galactic environment. In NGC~5194, the clusters located $\leq$1~$R_{\rm GMC}$ exhibit median ages of $\simeq$4~Myr, rising to $\simeq$6 and $\simeq$15~Myr when located in annuli between 1--2 and 2--3~$R_{\rm GMC}$, respectively. Star clusters unassociated with any GMC ($>$3~$R_{\rm GMC}$) exhibit ages of 40~Myr, 10~Myr older than the global median. 

\begin{figure}
\includegraphics[scale=0.43]{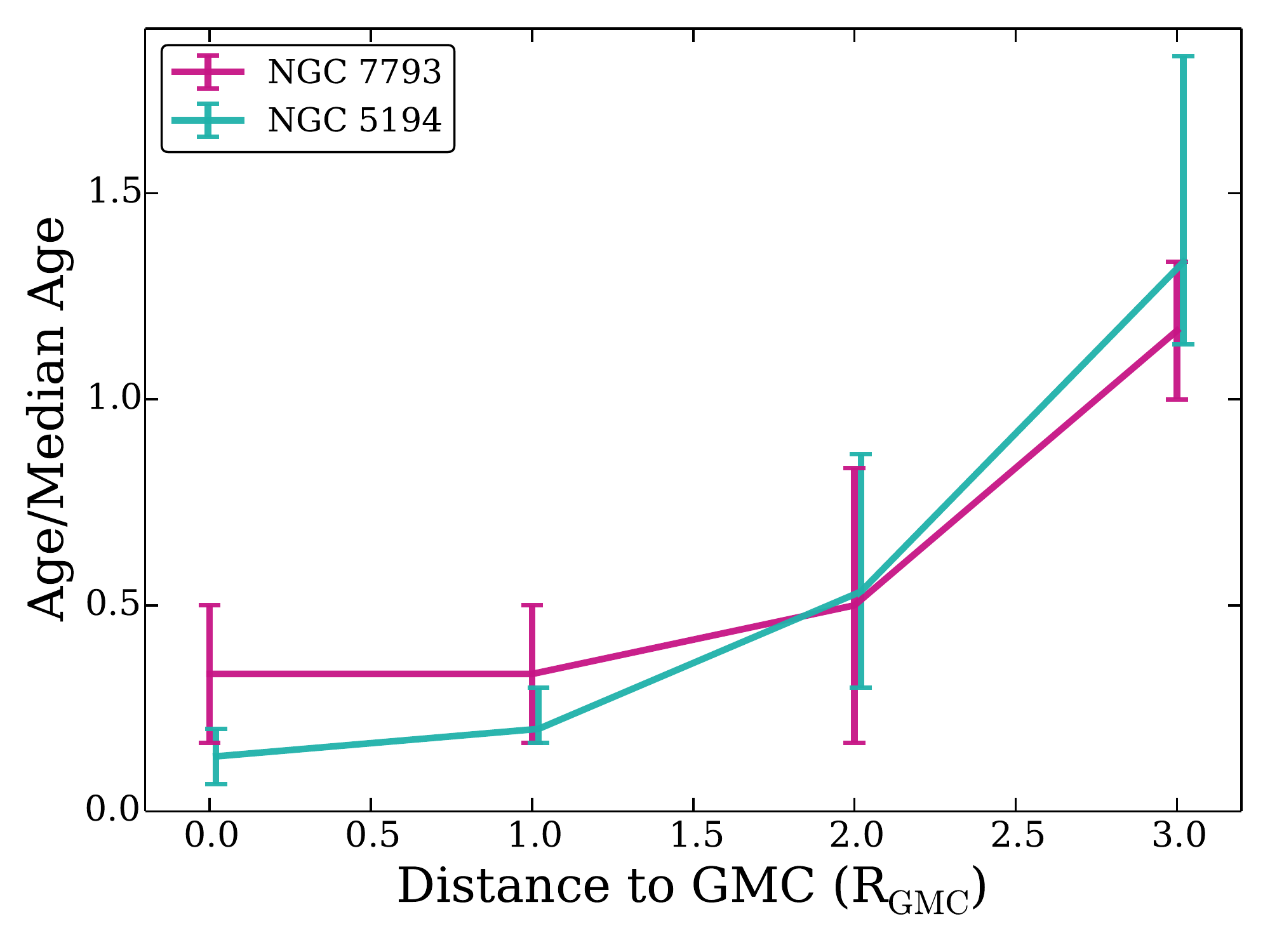}
\caption{
Relative age of the star cluster populations normalized by the median age of the global cluster populations for NGC~7793 (pink; Table~\ref{tab:1}) and NGC~5194 \citep[teal;][]{grasha18} as a function of distance from their closest GMC in units of radius of the GMC. Despite the different cluster properties and different scale lengths in the two galaxies, it takes half the median age of the star clusters to clear out the molecular material within 2$R_{\rm GMC}$. The slope gives the relative rate of clearing away the cloud material by the star clusters. The shallower slope for NGC~7793 implies that those the star clusters are capable of clearing out the molecular material faster (in relative units), resulting in a younger age for the star clusters to disassociate from their GMCs compared to NGC~5194. 
\label{fig:compare}}
\end{figure}

The difference in the association timescale between star clusters and GMCs, 2~Myr in NGC~7793 versus 4~Myr in NGC~5194, and unassociation cluster ages, 7 versus 40~Myr, can arise from the inherently different cluster, GMC, and ISM properties between the two galaxies. If we reconsider the age of clusters still associated with GMCs in the two galaxies as percentages of the global median age, by the time the star clusters have traveled a minimum of $2 R_{\rm GMC}$, they are half the median age of the total population in their respective systems. This is demonstrated in Figure~\ref{fig:compare}, where we visually demonstrate the relative age of the star clusters in both NGC~7793 and NGC~5194 as a function of distance from their GMC. Comparing in relative units allows us to compare the star cluster and GMC association in these two galaxies with different scales, cluster destruction rates, and median age of the populations. 

The slope of Figure~\ref{fig:compare} indicates the relative rate of how fast the star clusters clear away their natal molecular gas. The slope for NGC~7793 is less than that of NGC~5194, indicating that the star clusters in NGC~7793 are able to push away and destroy the gas faster (in relative units) than the star clusters in NGC~5194 are able to. Comparing these two galaxies, it is easier for the star clusters in NGC~7793 to separate from their molecular clouds. This faster timescale, resulting in a lower age for associated star clusters, can be the result of the lower surface brightness and lower pressure environment for this system compared to that of NGC~5194. 

This potentially informs on the impact of environmental stress forces on destroying young star clusters as the direct result of tidal forces with its local cloud \citep{elmegreenhunter10}. Star clusters are born at the densest region of the stellar hierarchy, slowly drifting away with time. As they age, the star clusters will travel from their initially structured regions of high-density with large tidal forces and collision rates towards low-density regions where the tidal forces and collision rates are smaller. While the time and length scales varies between the two galaxies, the pushing and destruction action on the clouds by the star clusters, however long and whatever distance that takes, can be similar in different environments when scaled to the mean cluster age and the mean cloud size. The difference in the median age of the star cluster population as a function of separation from a GMC in Figure~\ref{fig:compare} between NGC~7793 and NGC~5194 is significant at the 3.2--$\sigma$ level. Improvement can be made with futures studies that investigate this relation and its dependency with various galactic environments, such as ISM pressure, over a larger sample of galaxies.
%

\subsubsection{The Magellanic Clouds}
Within the SMC, \citet{mizuno01} found a positive correlation between young HII emission objects that trace star formation timescales of $\la$6~Myr and CO clouds, where $\sim$35\% of their sample of young emission objects are found to be associated with molecular clouds. Older, emissionless stellar objects (ages $\sim$6--100~Myr) did not show a strong correlation, with only $\sim14$\% of this older sample showing any type of spatial association with a molecular cloud. This suggests that CO clouds quickly dissipate after the formation of their star clusters and any correlations with older clusters can be the result of chance alignment. This temporal timescale for the spatial association of young star clusters with CO clouds is in agreement with results in the LMC with disassociation timescales of $\la$10~Myr \citep{fukui99, yamaguchi01}. Within the LMC, \citet{kawamura09} finds a disassociation timescale between clusters and GMCs of $\sim$7--10~Myr and additionally there is no significant observed difference in the size or line width between the varying evolutionary stages of GMCs. We derive a timescale of 2--3~Myr for star clusters to disassociate from GMCs in NGC~7793,  smaller than those measured in both the NGC~5194, the LMC, and the SMC.

\subsubsection{The Triangulum Galaxy M33}
The evolution of the molecular clouds in M33 has been extensively studied due to its proximity (840~kpc) and relatively low inclination. This provides an additional study to compare the star cluster and GMC conditions of NGC~7793 with the moderately interacting M33 system. \citet{onodera12} observed 100~pc resolution CO(3--2) with the Atacama Submillimeter Telescope Experiment. Using the same GMC catalog combined with optical data to identify young stellar groups, \citet{miura12} derived timescales for different stages in the evolutionary sequence of GMCs. They estimate that during the stage when GMCs are associated with young stellar groupings to last 5--10~Myr, consistent with the timescales we derive for NGC~7793. Due to the enhanced amount of dense gas found around the star forming regions, the star clusters do not destroy the GMCs but instead star formation propagates sequentially throughout the cloud until the gas is exhausted, with estimated lifetimes of 20--40~Myr for GMCs more massive than $\sim$10$^5$~\Msol. 

In a similar study to investigate the association between GMCs and star clusters, \citet{corbelli17} combine young and embedded star cluster candidates from the \citet{sharma11} Spitzer-24~\micron\ catalog with CO(2--1) IRAM observations with 49~pc resolution. They find an enhanced spatial correlation with separations of 17~pc between the star cluster candidates and GMCs. This separation is smaller than the separation we find in our survey of $\sim$40~pc, which can arise from an inherently different scale length in M33 (see Section \ref{sec:5194}) or can result from their star clusters identified with mid-IR emission peaks, preferentially selecting star clusters that more embedded and closer to their natal GMCs in comparison to our UV/optically selected star cluster sample. They recover a timescale of 5~Myr for their star cluster candidates to transition from the embedded to non-embedded phase, similar to the timescale we measure.

\citet{gonzalez12} combines the \citet{sharma11} star cluster catalog with neutral and molecular gas observations and find a strong trend for the maximum star cluster mass with increasing galactocentric radius as well as the gas surface density. The suppression of massive star clusters with increasing galactocentric radius and thus higher gas surface densities was also found by \citet{pflamm13}. This suggests that the pressure in the interstellar medium plays an important role in the formation of a star cluster by influencing the maximum mass of the pre-cluster core. As shown in Section~\ref{sec:5194}, higher ambient pressure also may lead to longer timescales for star cluster feedback to erode their natal gas clouds.

\subsection{The Two-Point Correlation Function}\label{sec:2pcf}
We implement the angular two-point correlation function $\omega(\theta)$ to measure the magnitude of clustering as a function of projected distance between the star clusters. A detailed description of the formalism and methodology of the two-point correlation function as applied to star clusters within other LEGUS galaxies can be found in both \citet{grasha15} and \citet{grasha17a}. The correlation function provides a method to identify the sizes of star-forming regions as well as common age structures to derive the randomization timescale for star-forming hierarchies. Here we list the details necessary for the application to the GMCs within NGC~7793. We have previously computed the correlation function for the star clusters in NGC~7793 and will compare those results to the distribution we find for the GMCs within the ALMA coverage. 

The angular correlation function $\omega(\theta)$ is the probability of finding a neighboring object within an angular separation $\theta$ above what is expected for a Poisson distribution $\mathrm{d}P = N^2 [ 1 +  \omega(\theta)]\ \mathrm{d}\Omega_1 \mathrm{d}\Omega_2$, where $N$ is the surface density of clusters per steradian with two infinitesimal elements of solid angle $\mathrm{d}\Omega_1$ and $\mathrm{d}\Omega_2$ \citep{peebles80}. A clustered distribution has an excess of pair counts at small separations, resulting in a declining power law distribution of $1 + \omega(\theta)$ toward larger scale lengths. 

We estimate the correlation function using the \citet{landy93} estimator by counting pairs of star clusters (or GMCs) in increasing annuli and comparing those cluster counts to expectations from unclustered distributions. We supplement the GMC data with a catalog of sources that are randomly populated with the same sky coverage as the ALMA observations. The correlation function of stellar components has been demonstrated to be well-described with a power law $1+\omega(\theta) = A_{\omega}\theta^{\alpha}$ following the convention of \citet{calzetti89}, where the slope $\alpha$ measures the strength of the clustering and the amplitude $A_{\omega}$ measures the correlation length of the clustering when $\theta>1$. Interstellar gas exhibits a hierarchical morphology structure with a typical fractal dimension of $D2 = \alpha+2 \simeq 1.5$ \citep{elmegreen06}, allowing the projected correlation function to serve as an approximate comparison for the distribution of individual GMCs and star clusters to the expected hierarchical distribution of the ISM. 

\begin{figure}
\includegraphics[scale=0.43]{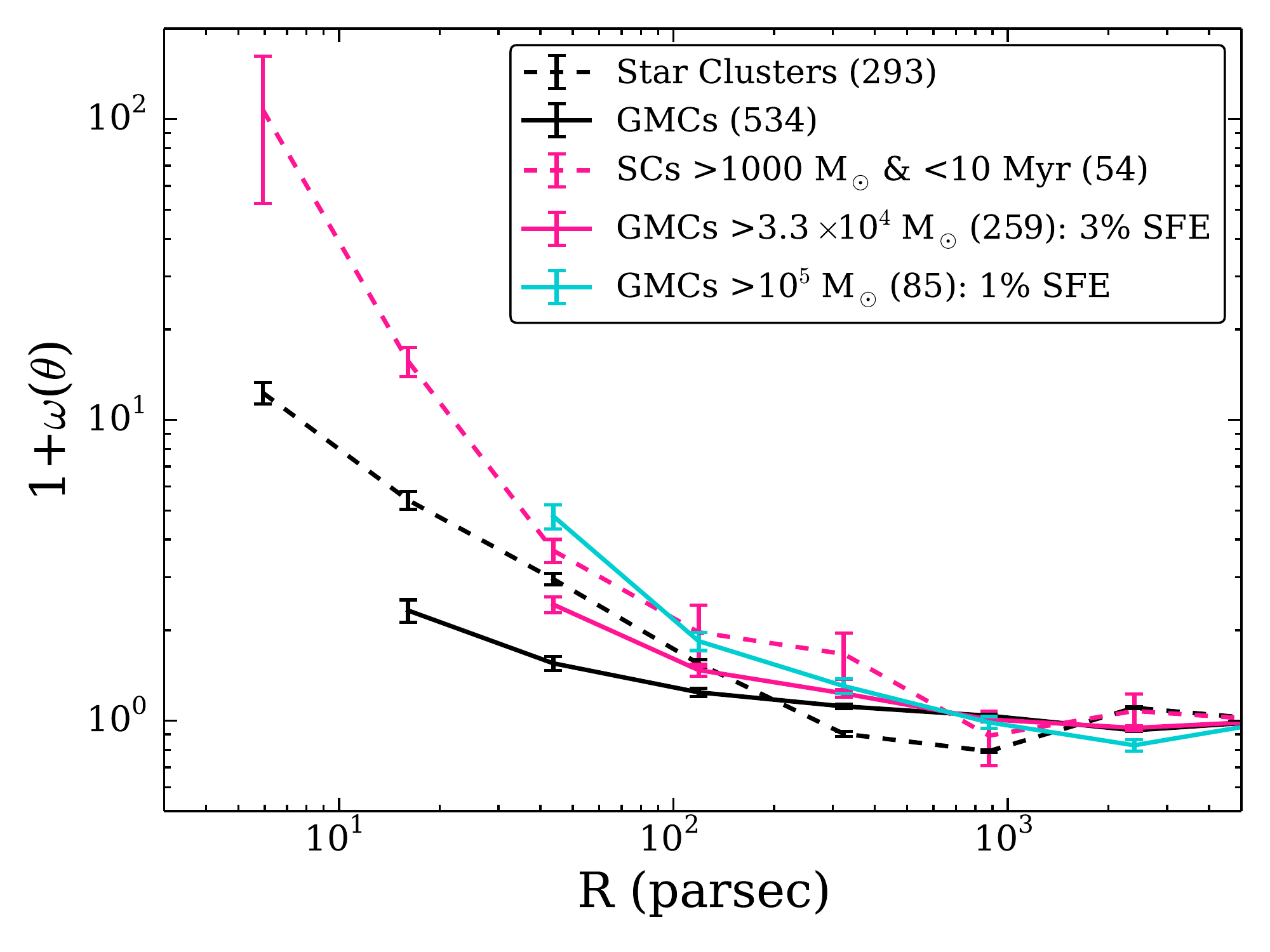}
\caption{
Two-point correlation function $1+\omega(\theta)$ for the star clusters (dotted lines) and the GMCs (solid lines). The black line shows the global average for both the star clusters and the GMCs that are within the ALMA coverage (gray box of Figure~\ref{fig:7793}). The numbers in parentheses list the number of objects. The power law relation is expected for a hierarchical distribution and is consistent with the distribution of star clusters and stars within other galaxies. The observed clustering is very flat for the GMCs given the strong age-dependency to the clustering. To compare how GMCs and star clusters respond to clustering, we show the distribution of the most massive and youngest star clusters to that of the most massive GMCs. We show the star clusters with a mass cut at 1000~\Msol (dashed pink line) and the equivalent GMC distribution with mass cuts assuming a 3\% SFE (solid pink line) and a 1\% SFE (solid teal line). The clustering in the distribution of the GMCs increase with mass, becoming similar to the distribution of the youngest and most massive star clusters. The power law fits are performed over the scale lengths 40--800~pc to aid in the comparsion between the different samples. 
\label{fig:2pcf}}
\end{figure}

Figure~\ref{fig:2pcf} shows the two-point correlation function for the GMCs and the subset of clusters located within the ALMA coverage. The random catalog follows the outline of the ALMA coverage in Figure~\ref{fig:7793}. The star clusters exhibit a smooth and steady decline with increasing radius, well described with a power law which is nearly identical to the global star cluster distribution \citep[see][]{grasha17a}. We perform single power law fits in log-log space to both distributions over scale lengths 40--800~pc when $1+\omega(\theta)>1$. We recover a slope of $-0.35\pm0.02$ for the star clusters whereas the GMCs exhibit a considerably shallower slope of $-0.18\pm0.02$.

The distribution of the GMCs is significantly flatter than the star cluster distribution, a result also seen in NGC~5194 \citep{grasha18}. The close association in time between GMCs and star formation suggests that the hierarchy of star clusters should be reflected by the GMCs \citep[see, e.g.,][]{dobbs14}, however, mirroring the exact spatial distribution would indicate that each GMC can result in the creation of only one star cluster. The observed excess in the clustering of the star clusters compared to the distribution of the GMCs in NGC~7793 and NGC~5194 strongly suggests that the natal distribution of star clusters must be more structured than that of the GMCs. This could arise from requiring a GMC to produce more than one star cluster, though the production of star clusters does not need to be simultaneous but could be sequential, and/or indicates that not all individual GMCs form a star cluster. Molecular clouds do evolve over time and exhibit different levels of star formation activity \citep[e.g.,][]{yamaguchi01, kawamura09, ochsendorf17}. In particular, \citet{kawamura09} found that not all GMCs show evidence of star formation and more evolved clouds appear to be associated with optical stellar clusters, and thus, we may be sampling star formation regions in different evolutionary states where exposed star clusters reside in regions where the molecular gas reservoir has already disrupted.

Not all GMCs will form a star cluster; the inclusion of the entire GMC population in the calculation for Figure~\ref{fig:2pcf} may quite possibly be erasing the clustered distribution of the GMCs that are most likely forming the star clusters. To more fairly compare the distributions of the star clusters and GMCs, we limit the star clusters to those more massive than 1000 \Msol\ and match the mass limit between the star clusters and GMCs by assuming a star formation efficiency (SFE) of 1\% and 3\%. This allows us to exclude low mass GMCs that are not potentially actively undergoing star formation and not forming the current stellar populations in our catalogs. 

There is a general increase in the slope of the GMC distribution with increasing mass although we are unable to compare the distribution of the massive GMCs below scales of $\sim$40~pc due to their small numbers. The increased clustering for increasing GMC mass implies that our optically identified star clusters arise from a specific subset of molecular clouds. We conclude that it is the most massive clouds that are more likely to produce the star clusters we observe. 

Despite the similar spatial distributions between the most massive GMCs to the youngest and most massive star clusters, the clustering observed in the GMCs is still weaker than that observed for the star clusters. Table~\ref{tab:slopes} lists the slopes for the different correlation functions in Figure~\ref{fig:2pcf}. The parameters reported in Table~\ref{tab:slopes} are only calculated in the range of 40--800 pc to compare the same spatial scales between the star clusters and GMCs. 

Star clusters originate in groupings within clustered GMCs and a stronger excess of close neighbors increases when considering the youngest clusters of the hierarchy \citep{grasha17a}. The different lifetimes of star clusters and GMCs may impact the dispersal timescale from the structures, and hence, the survival of the large scale hierarchies. The cluster age distribution for NGC~7793 suggests the presence of disruption \citep{mora09, silvavilla11}. Thus the cluster dissolution timescale will have some degree of impact on the clustering results as the estimated randomization timescale for the dispersion of the hierarchies is fairly short, $\sim$40--60~Myr for NGC~7793 \citep{grasha17a}. On the other hand, the lifetimes of typical GMCs, upwards of tens of~Myr depending on the system \citep{murray11, heyer15, meidt15}, are significantly shorter than the derived lifetimes of star-forming complexes. As a result, the hierarchical structure present in the spatial distribution of the GMCs disappears even quicker due to their rapid destruction. 

A steeper star cluster distribution compared to the GMC distribution could be the result of the formation of a star cluster triggering the additional formation of other clusters within a cloud \citep{elmegreen77}. The shallower slope of GMCs may be further enhanced by the CO observations not being sensitive to the dense peaks of CO-dark molecular gas, making it difficult to detect the structures of the dense ISM where the vast majority of H$_2$ may be located and where stars are actively forming \citep{grenier05, glover16}. 
%
\begin{table}
 \caption{Correlation Function power law fits. Columns list the
(1) Age or mass range for the computation of the two-point correlation function; 
(2) Number of sources; 
(3) Amplitude $A$ of the correlation function in the range 40--800 pc; and
(4) Slope $\alpha$ of the correlation function in the range 40--800 pc.  }
 \label{tab:slopes}
 \begin{tabular}{lccc}
  \hline
\multicolumn{4}{c}{Star Clusters}\\
\hline
All & 293 & 8.3$\pm$0.2 & $-$0.35$\pm$0.02  \\
Mass $\geq$1000 \Msol\ \& Age $\leq$10 Myr  & 54 & 20$\pm$3 & $-$0.45$\pm$0.06  \\
\hline\\
\multicolumn{4}{c}{Molecular Clouds}\\
\hline
All & 534 & 3.2$\pm$0.2 & $-$0.18$\pm$0.02   \\
Mass $\geq$3.3$\times$10$^4$~\Msol\ & 259 & 5.2$\pm$0.2 & $-$0.25$\pm$0.02  \\
Mass $\geq$1$\times$10$^5$~\Msol\ & 85 & 18.6$\pm$1.4 & $-$0.44$\pm$0.03  \\
  \hline
 \end{tabular}
\end{table}

\section{Summary and Conclusion}\label{sec:summary}
The impact of spiral structure and feedback from stellar populations on molecular clouds has broad-ranging implications for star formation in the local universe. The high-quality homogeneous star cluster catalogs from the HST LEGUS project \citep{calzetti15a} with reliable age measurements down to a few Myr, combined with exquisite molecular CO data, are both essential in addressing the relationship of star clusters with the properties of molecular gas in a consistent and precise manner. In this work, we present a study on the connection between the star clusters in the flocculent LEGUS galaxy NGC~7793 with high-resolution molecular gas from ALMA at $\sim$20~pc, resolving down to individual GMCs. We compare the locations of star clusters to that of GMCs to investigate the age at which star clusters remain associated with their molecular gas. 

Our main results are summarized as follows.
\begin{enumerate}
	\item The youngest star clusters are predominately located in close vicinity to a GMC, showing an enhanced spatial correlation between the molecular clouds and young stellar clusters ($\la$10~Myr; Figure~\ref{fig:sc_gmc_mindistance}) on scales $\la$40~pc. 
	\item Star clusters that are still associated with a GMC (i.e., located within the footprint of a GMC) exhibit median ages of 2~Myr compared to the global median age of the clusters at 6~Myr. Star clusters that reside at distances $>3 R_{\rm GMC}$ to the nearest GMC are considered unassociated with any molecular cloud and exhibit median ages of 7~Myr (Figure~\ref{fig:dist_agemass}). These age measurements help estimate the erosion of molecular gas from the young star clusters, helping to constrain the association timescale of star clusters and GMCs environments with recent star formation. We conclude that star clusters disassociate from their natal clouds in very short timescales of 2--3~Myr. 
	\item In comparison to a similar study of NGC~5194 \citep{grasha18}, it takes half of the median age of the entire cluster population to clear out the molecular material and separate more than 2$R_{\rm GMC}$. This finding is intriguing given the drastically different timescales, distances, cluster and ISM properties between the two galaxies. Furthermore, the star clusters of NGC~7793 appear to clear out the immediate molecular material faster than the star clusters in NGC~5194 (Figure~\ref{fig:compare}), giving rise to the observed timescale of 2--3~Myr for star clusters to disassociate from their molecular clouds in NGC~7793. The environmental stress of the interactions between star clusters and their host clouds may very well impact the destruction of young star clusters. 
	\item We implement the two-point correlation function to compare the hierarchical distribution of the GMCs to that of the star clusters. The star clusters are well described by a decreasing power law with increasing spatial scale, with a slope of $-0.35\pm0.02$ over the range 40--800~pc. The GMCs show a significantly flatter distribution with a power law slope of $-0.25\pm0.02$ over the same scale lengths, thus star clusters show a stronger clustering than GMCs. This suggests that not all GMCs form a star cluster and those that do most likely form more than one star cluster. 	
	\item The clustering observed for the star clusters and GMCs become similar to each other when comparing the most massive GMCs to the youngest and most massive star clusters. Over the range 40--800~pc, star clusters younger than 10 Myr and more massive than 1000~\Msol\ show a slope of $-0.45\pm0.06$ and GMCs with masses greater than 10$^5$~\Msol\ an identical slope of $-0.44\pm0.03$. This shows the importance of matching the mass limit for star clusters and GMCs by assuming a SFE of a few percent. This allows for a better identification of the subset of GMCs that are responsible for the formation of the current population of star clusters. Only after we place these masscuts do we fairly compare the clustered distribution of the star clusters to the GMCs they formed from and derive similar fractal distributions. 
\end{enumerate}

In the future, we aim to broaden the method implemented here to a larger range of galactic environments to determine the exact effect of the local environment on the disassociation timescale between star clusters and GMCs. We also seek to further improve the derivation of the disassociation timescales at different GMC evolutionary stages, by expanding our sample to include IR selected, dusty star clusters. Finally, we plan to explore how properties of the GMCs may affect the properties of the star clusters they produce.

\section*{Acknowledgements}
We are grateful for the valuable comments on this work by an anonymous referee that improved the scientific outcome and quality of the paper.
Based on observations made with the NASA/ESA Hubble Space Telescope, obtained at the Space Telescope Science Institute, which is operated by the Association of Universities for Research in Astronomy, under NASA Contract NAS 5--26555. These observations are associated with Program 13364 (LEGUS). Support for Program 13364 was provided by NASA through a grant from the Space Telescope Science Institute. 
This research has made use of the NASA/IPAC Extragalactic Database (NED) which is operated by the Jet Propulsion Laboratory, California Institute of Technology, under contract with NASA. This paper makes use of the following ALMA data: ADS/JAO.ALMA \#2015.1.00782.S. ALMA is a partnership of ESO (representing its member states), NSF (USA) and NINS (Japan), together with NRC (Canada) and NSC and ASIAA (Taiwan) and KASI (Republic of Korea), in cooperation with the Republic of Chile. The Joint ALMA Observatory is operated by ESO, AUI/NRAO and NAOJ. The National Radio Astronomy Observatory is a facility of the National Science Foundation operated under cooperative agreement by Associated Universities, Inc. 
Parts of this research were supported by the Australian Research Council Centre of Excellence for All Sky Astrophysics in 3 Dimensions (ASTRO 3D), through project number CE170100013.
A.A. acknowledges the support of the Swedish Research Council (Vetenskapsr\r{å}det) and the Swedish National Space Board (SNSB). MF acknowledges support by the Science and Technology Facilities Council [grant number ST/P000541/1]. This project has received funding from the European Research Council (ERC) under the European Union's Horizon 2020 research and innovation programme (grant agreement No 757535).

\textit{Software}:
Astropy \citep{astropy}, 
iPython \citep{ipython}, 
Matplotlib \citep{matplotlib}, 
Numpy \citep{numpy}, 
TreeCorr \citep{treecorr}



\begin{thebibliography}{}
{\footnotesize
\bibitem[Adamo \etal(2010)]{adamo10} Adamo, A., \"{O}stlin, G., Zackrisson, E., \etal\ 2010, \mnras, 407, 870
\bibitem[Adamo \etal(2012)]{adamo12} Adamo, A., Smith, L.J., Gallagher, J.S., \etal\ 2012, \mnras, 426, 1185
\bibitem[Adamo \etal(2017)]{adamo17} Adamo, A., Ryon, J.E., Messa, M., \etal\ 2017, \apj, 841, 131
\bibitem[Andrews \etal(2013)]{andrews13} Andrews, J.E., Calzetti, D. Chandar, R., \etal\ 2013, \apj, 767, 51
\bibitem[Andrews \etal(2014)]{andrews14} Andrews, J.E., Calzetti, D. Chandar, R., \etal\ 2014, \apj, 793, 4
\bibitem[Ashworth \etal(2017)]{ashworth17} Ashworth, G., Fumagalli, M., Krumholz, M.R., \etal\ 2017, \mnras, 469, 2464
\bibitem[Astropy(2013)]{astropy} Astropy Collaboration \etal\ 2013, \aap, 558, A33
\bibitem[Bastian \etal(2005)]{bastian05} Bastian, N., Gieles, M., Lamers, H.J.G.L.M., Scheepmaker, R.A., \& de Grijs, R. 2005, \aap, 431, 905 
\bibitem[Battersby \etal(2017)]{battersby17} Battersby, C., Bally, J., Svoboda, B. 2017, \apj, 835, 263 
\bibitem[Bertin \& Arnouts(1996)]{bertin96} Bertin, E. \& Arnouts, S. 1996, \aaps, 117, 393
\bibitem[Calzetti \etal(1989)]{calzetti89} Calzetti, D., Giavalisco, M., \& Ruffini, R. 1989, \aap, 226, 1
\bibitem[Calzetti \etal(2000)]{calzetti00} Calzetti, D., Armus, L., Bohlin, R.C., \etal\ 2000, \apj, 533, 682
\bibitem[Calzetti \etal(2010)]{calzetti10} Calzetti, D., Chandar, R. Lee, J.C., \etal\ 2010, \apj, 719, 158
\bibitem[Calzetti \etal(2015)]{calzetti15a} Calzetti, D., Lee, J.C., Sabbi, E. \etal\ 2015, \aj, 149, 51
\bibitem[Carignan \& Puche(1990)]{carignan90} Carignan, C. \& Puche, D. 1990, \aj, 100, 394
\bibitem[Cervi\~{n}o \etal(2002)]{cervino02} Cervi\~{n}o, M. Valls-Gabaud, D, Luridiana, V., \& Mas-Hesse, J.M. 2002, \aap, 381, 51
\bibitem[Colombo \etal(2014)]{colombo14a} Colombo, D., Hughes, A., Schinnerer, E., \etal\ 2014, \apj, 784, 3 
\bibitem[Corbelli \etal(2017)]{corbelli17} Corbelli, E., Braine, J., Bandiera, R., \etal\ 2017, \aap, 601, 146
\bibitem[da Silva \etal(2012)]{dasilva12} da Silva, R.L., Fumagalli, M., \& Krumholz, M. 2012, \apj, 745, 145
\bibitem[Dicaire \etal(2008)]{dicaire08} Dicaire, I., Carignan, C., Amram, P., \etal\ 2008, \aj, 135, 2038 
\bibitem[Dobbs \& Pringle(2013)]{dobbs13} Dobbs, C.L. \& Pringle, J.E. 2013, \mnras, 432, 653
\bibitem[Dobbs \etal(2014)]{dobbs14} Dobbs, C.L., Pringle, J.E., \& Naylor, T. 2014, \mnras, 437, L31
\bibitem[Efremov \& Elmegreen(1998)]{efremov98} Efremov, Y.N. \& Elmegreen, B.G. 1998, \mnras, 299, 588
\bibitem[Elmegreen \& Lada(1977)]{elmegreen77} Elemgreen, B.G. \& Lada, C.J. 1977, \apj, 214, 725
\bibitem[Elmegreen \& Falgarone(1996)]{elmegreenfalgarone96} Elmegreen, B.G. \& Falgarone, E. 1996, \apj, 471, 816 
\bibitem[Elmegreen \& Efremov(1996)]{elmegreen96} Elmegreen, B.G. \& Efremov, Y.N. 1996, \apj, 466, 802
\bibitem[Elmegreen \& Efremov(1997)]{elmegreen97} Elmegreen, B.G. \& Efremov, Y.N. 1997, \apj, 480, 235
\bibitem[Elmegreen \etal(2006)]{elmegreen06} Elmegreen, B.G., Elmegreen, D.M., Chandar, R., \etal\ 2006, \apj, 644, 879
\bibitem[Elmegreen \& Hunter(2010)]{elmegreenhunter10} Elmegreen, B.G. \& Hunter, D.A. 2010, \apj, 712, 604 
\bibitem[Elmegreen \etal(2014)]{elmegreen14} Elmegreen, D.M., Elmegreen, B.G., Adamo, A., \etal\ 2014, \apjl, 787, L15
\bibitem[Ferland \etal(1998)] {ferland98} Ferland, G.J., Korista, K.T., Verner, D.A., Ferguson, J.W., Kingdon, J.B., \& Verner, E.M. 1998, \pasp, 110, 761
\bibitem[Ferland \etal(2013)]{ferland13} Ferland, G.J., Porter, R.L., van Hoof, P.A.M., \etal\ 2013, Rev. Mexicana Astron. Astrofis., 49, 137
\bibitem[Fukui \etal(1999)]{fukui99} Fukui, Y., Mizuno, N., Yamaguchi, R., Mizuno, A., Onishi, T., \etal\ 1999, \pasj, 51, 745
\bibitem[Glover \& Smith(2016)]{glover16} Glover, S.C.O. \& Smith, R.J. 2016, \mnras, 462, 3011
\bibitem[Gomez \etal(1993)]{gomez93} Gomez, M., Hartmann, L., Kenyon, S.J., \& Hewett, R. 1993, \aj, 105, 1927
\bibitem[Gouliermis \etal(2014)]{gouliermis14} Gouliermis, D.A., Hony, S., \& Klessen, R.S. 2014, \mnras, 439, 3775
\bibitem[Gouliermis \etal(2015)]{gouliermis15} Gouliermis, D.A., Thilker, D., Elmegreen, B.G., \etal\ 2015, \mnras, 452, 3508
\bibitem[Gouliermis \etal(2017)]{gouliermis17} Gouliermis, D.A., Elmegreen, B.G., Elmegreen, D.M., \etal\ 2017, \mnras, 468, 509
\bibitem[Gonz\'{a}lez-L\'{o}pezlira \etal(2012)]{gonzalez12} Gonz\'{a}lez-L\'{o}pezlira, R.A., Pflamm-Altenburg, J., \& Kroupa, P. 2012, \apj, 761, 124
\bibitem[Grasha \etal(2015)]{grasha15} Grasha, K., Calzetti, D., Adamo, A., \etal\ 2015, \apj, 815, 93
\bibitem[Grasha \etal(2017a)]{grasha17a} Grasha, K., Calzetti, D., Adamo, A., \etal\ 2017a, \apj, 840, 113
\bibitem[Grasha \etal(2017b)]{grasha17b} Grasha, K., Elmegreen, B.G., Calzetti, D., \etal\ 2017b, \apj, 842, 25
\bibitem[Grasha \etal(2018)]{grasha18} Grasha, K., Calzetti, D., Fedorkeno, K., \etal\ 2018, \apj, submitted
\bibitem[Grenier \etal(2005)]{grenier05} Grenier, I.A., Casandjian, J.-M., Terrier, R. 2005, Science, 307, 1292
\bibitem[Heyer \& Dame(2015)]{heyer15} Heyer, M. \& Dame, T.M. 2015, \araa, 53, 583
\bibitem[Hopkins(2013a)]{hopkins13a} Hopkins, P.F. 2013a, \mnras, 428, 1950
\bibitem[Hopkins(2013b)]{hopkins13b} Hopkins, P.F. 2013b, \mnras, 430, 1653
\bibitem[Hughes \etal(2013)]{hughes13} Hughes, A., Meidt, S., Schinnerer, E., et al. 2013, \apj, 779, 44
\bibitem[Hunter(2007)]{matplotlib} Hunter J.D. 2007, Comput. Sci. Eng., 9, 90 
\bibitem[Kahre \etal(2018)]{kahre18} Kahre, L., Walterbos, R.A., Kim, H., \etal\ 2018, \apj, 855, 133
\bibitem[Jarvis \etal(2004)]{treecorr} Jarvis, M., Bernstein, G., \& Jain, B. 2004, \mnras, 352, 338 
\bibitem[Johnstone \etal(2000)]{johnstone00} Johnstone, D., Wilson, C.D., Moriarty-Schieven, G., Joncas, G., Smith, G., Gregersen, E., \& Fich, M. 2000, \apj, 545, 327
\bibitem[Johnstone \etal(2001)]{johnstone01} Johnstone, D., Fich, M., Mitchell, G.F., \& Moriarty-Schieven, G. 2001, \apj, 559, 307
\bibitem[Kawamura \etal(2009)]{kawamura09} Kawamura, A., Mizuno, Y.,  Minamidani, T., \etal\ 2009, \apjs, 184, 1
\bibitem[Kirk \& Meyers(2012)]{kirk12} Kirk, H. \& Meyers, P.C. 2012, \apj, 745, 131
\bibitem[Kreckel \etal(2018)]{kreckel18} Kreckel, K., Faesi, C., Kruijssen, J.M.D., \etal\ 2018, \apjl, in press (arxiv: 1807.11506)
\bibitem[Kroupa(2001)]{kroupa01} Kroupa, P. 2001, \mnras, 322, 231
\bibitem[Krumholz \etal(2015a)]{krumholz15a} Krumholz, M.R., Adamo, A., Fumagalli, M., \etal\ 2015a, \apj, 812, 147
\bibitem[Krumholz \etal(2015b)]{krumholz15b} Krumholz, M.R., Fumagalli, M., da Silva, R.L., Rendahl, T., \& Parra, J. 2015b, \mnras, 452, 1447
\bibitem[Lada(1987)]{lada87} Lada, C.J. 1987, IAUS, 115, 1
\bibitem[Landy \& Szalay(1993)]{landy93} Landy, S.D. \& Szalay, A.S. 1993, \apj, 412, 64
\bibitem[Leisawitz \etal(1989)]{leisawitz89} Leisawitz, D., Bash, F.N., \& Thaddeus, P. 1989, \apjs, 70, 731
\bibitem[Leitherer \etal(1999)]{leitherer99} Leitherer, C., Schaerer, D., Goldader, J.D., \etal\ 1999, \apjs, 123, 3
\bibitem[Marks \etal(2012)]{marks12} Marks, M., Kroupa, P., Dabringhausen, J., \& Pawlowski, M.S. 2012, \mnras, 422, 2246
\bibitem[Meidt \etal(2015)]{meidt15} Meidt, S.E., Hughes, A., Dobbs, C.L, \etal\ 2015, \apj, 806, 72
\bibitem[Messa \etal(2018a)]{messa18a} Messa, M., Adamo, A., \"{O}stlin, G., \etal\ 2018a, \mnras, 473, 996
\bibitem[Messa \etal(2018b)]{messa18b} Messa, M., Adamo, A., Calzetti, D., \etal\ 2018b, \mnras, 477, 1683
\bibitem[Matthews \etal(2018)]{matthews18} Matthews, A.M., Johnson, K.E., Whitmore, B.C., Brogan, C.L., Leroy, A.K., \& Indebetouw, R. 2018, ApJ, 862, 147
\bibitem[Miura \etal(2012)]{miura12} Miura, R.E., Kohno, K., Tosaki, T., Espada, D., Hwang, N., \etal\ 2012, \apjs, 761, 37
\bibitem[Mizuno \etal(2001)]{mizuno01} Mizuno, N., Tubio, M., Yamaguchi, R., Onishi, T., \& Fukui, Y. 2001, \pasj, 53, L45
\bibitem[Mora \etal(2009)]{mora09} Mora, M.D., Larsen, S.S., Kissler-Patig, M., Brodie, J.P., \& Richtler, T. 2009, \aap, 501, 949
\bibitem[Muraoka \etal(2016)]{muraoka16} Muraoka, K., Takeda, M., Yanagitani, K., \etal\ 2016, \pasp, 68, 18
\bibitem[Murray(2011)]{murray11} Murray, N. 2011, \apj, 729, 133
\bibitem[Odekon(2008)]{odekon08} Odekon, M.C. 2008, \apj, 681, 1248
\bibitem[Ochsendorf \etal(2017)]{ochsendorf17} Ochsendorf, B.B., Meixner, M., Roman-Duval, J., Rahman, M., \& Evans II, N.J. 2017, \apj, 841, 109
\bibitem[Onodera \etal(2012)]{onodera12} Onodera, S., Kuno, N., Tosaki, T., \etal\ 2016, \pasj, 64, 133
\bibitem[Padoan \& Nordlund(2002)]{padoan02} Padoan, P. \& Nordlund, \r{A}. 2002, \apj, 576, 870
\bibitem[Peebles(1980)]{peebles80} Peebles, P.J.E. 1980, `The Large-Scale Structure of the Universe' (Princeton, N.J.: Princeton University Press)
\bibitem[P\'{e}rez \& Granger(2007)]{ipython} P\'{e}rez, F. \& Granger, B.E. 2007, Comput. Sci. Eng., 9, 21 
\bibitem[Pety \etal(2013)]{pety13} Pety, J., Schinnerer, E., Leroy, A.K., \etal\ 2013, \apj, 779, 43
\bibitem[Pflamm-Altenburg \etal(2013)]{pflamm13} Pflamm-Altenburg, J., Gonz\'{a}lez-L\'{o}pezlira, R.A., \& Kroupa, P. 2013, \mnras, 435, 2604
\bibitem[Pietrzy\'{n}ski \etal(2010)]{pietrzynski10} Pietrzy\'{n}ski, G., Gieren, W., Hamuy, M., \etal\ 2010, \aj, 140, 1475
\bibitem[Pilyugin \etal(2014)]{pilyugin14} Pilyugin, L.S., Grebel, E.K., \& Kniazev, A.Y. 2014, \aj, 147, 131
\bibitem[Ram\'{i}rez Alegr\'{i}a \etal(2016)]{ramirez16} Ram\'{i}rez Alegr\'{i}a, S., Borissova, J., Chen\'{e}, A.-N., \etal\ 2016, \aap, 588, 40
\bibitem[Renaud \etal(2013)]{renaud13} Renaud, F., Bournaud, F., Emsellem, E., \etal\ 2013, \mnras, 436, 1836
\bibitem[Rosolowsky \& Leroy(2006)]{rosolowsky06} Rosolowsky, E. \& Leroy, A.K. 2006, \pasp, 118, 590
\bibitem[Ryon \etal(2017)]{ryon17} Ryon, J.E., Gallagher, J.S., Smith, L.J., \etal\ 2017, \apj, 841, 92
\bibitem[S\'{a}nchez \etal(2010)]{sanchez10} S\'{a}nchez, N., A\~{n}ez, N., Alfaro, E., Odekon, M.C. 2010, \apj, 720, 541
\bibitem[Schinnerer \etal(2013)]{schinnerer13} Schinnerer, E., Meidt, S.E, Pety, J., \etal\ 2013, \apj, 779, 42
\bibitem[Schlafly \& Finkbeiner(2011)]{schlafly11} Schlafly, E.F. \& Finkbeiner, D.P. 2011, \apj, 737, 103
\bibitem[Sharma \etal(2011)]{sharma11} Sharma, S., Corbelli, E., Giovanardi, C., Hunt, L.K., \& Palla, F. 2011, \aap, 534, 96
\bibitem[Silva-Villa \& Larsen(2011)]{silvavilla11} Silva-Villa, E. \& Larsen, S.S. 2011, \aap, 529, 25
\bibitem[Stephens \etal(2017)]{stephens17} Stephens, I.W., Gouliermis, D.A., Looney, L.W., \etal\ 2017, \apj, 834, 94
\bibitem[Sun \etal(2017)]{sun17a} Sun, N.-C., de Grijs, R., Subramanian, S. \etal\ 2017, \apj, 835, 171
\bibitem[Sun \etal(2018)]{sun18} Sun, N.-C., de Grijs, R., Cioni, M.-R. L., \etal\ 2018, \apj, 858, 31
\bibitem[van der Walt \etal(2011)]{numpy} van der Walt, S., Colbert, S.C., \& Varoquaux, G. 2011, Comput. Sci. Eng., 13, 22 
\bibitem[V\'{a}zquez \& Leitherer(2005)]{vazquez05} V\'{a}zquez, G.A. \& Leitherer, C. 2005, \apj, 621, 695
\bibitem[Weidner \etal(2013)]{weidner13} Weidner, C., Kroupa, P., \& Pflamm-Altenburg, J. 2013, \mnras, 434, 84
\bibitem[Whitmore \etal(2014)]{whitmore14} Whitmore, B.C., Brogan, C., Chandar, R., \etal\ 2014, \apj, 795, 156
\bibitem[Yamaguchi \etal(2001)]{yamaguchi01} Yamaguchi, R., Mizuno, N., Mizuno, A., \etal\ 2001, \pasj, 53, 985
\bibitem[Zackrisson \etal(2011)]{zackrisson11} Zackrisson, E., Rydberg, C.-E., Schaerer, D., \"{O}stlin, G., \& Tuli, M. 2011, \apj, 740, 13
\bibitem[Zhang \etal(2001)]{zhang01} Zhang, Q., Fall, S.M., \& Whitmore, B.C. 2001, \apj, 561, 727
}
\end{thebibliography}


\bsp	
\label{lastpage}
\end{document}